\documentclass[12pt,preprint]{aastex} \tightenlines
\received{} \accepted{}

\def\ltsima{$\; \buildrel < \over \sim\;$}
\def\simlt{\lower.5ex\hbox{\ltsima}}            % < over MMM
\def\gtsima{$\; \buildrel > \over \sim \;$}
\def\simgt{\lower.5ex\hbox{\gtsima}}            % > over MMM

\def\m2{{M$_{\rm H_2}$\/}}

\begin{document}

\title{\sc A Multiwavelength study of NLS1s galaxies from the Second Byurakan Survey}

\author{\sc J. A. Stepanian\altaffilmark{1}, E. Ben{\'\i}tez\altaffilmark{1},
Y. Krongold\altaffilmark{2}, I. Cruz-Gonz{\'a}lez\altaffilmark{1},
J. A. de Diego\altaffilmark{1}, V. Chavushyan\altaffilmark{3}, R.
M{\'u}jica\altaffilmark{3}, D. Dultzin-Hacyan\altaffilmark{1}, T.
Verdugo\altaffilmark{1}}

\altaffiltext{1}{Instituto de Astronom{\'\i}a, UNAM, Apartado
Postal 70-264, 04510 M\'exico DF. e-mail:
jstep@astroscu.unam.mx,erika@astroscu.unam.mx,
irene@astroscu.unam.mx} \altaffiltext{2}{Harvard-Smithsonian
Center for Astrophysics, 60 Garden Street,
Cambridge,MA02138;e-mail: krongold@head-cfa.cfa.harvard.edu }
\altaffiltext{3}{Instituto Nacional de Astrof{\'\i}sica, Optica y
Electronica,INAOE,Apartado Postal 512 y 216, 7200, Puebla, Pue.,
M\'exico. e-mail: vahram@inaoep.mx, mujica@inaoep.mx}

\begin{abstract}

In this work we present a multiwavelength study of narrow-line
Seyfert 1 galaxies (NLS1s) discovered in the Second Byurakan
Survey (SBS). The sample consists of 26 objects, which have
$M_{B}\ge -23.0$, or $-19.9 > M_{B} > -23.0$, $0.0243 < z < 0.317$
and $15.2 < B < 19.0$. For these objects, we present accurate
coordinates, magnitudes, redshifts and the identification of
optical objects with X-ray, IR and radio sources. Several galaxies
are identified as X-ray or radio sources for the first time. We
also report spectroscopic and photometric data,  the spectral
energy distribution, $\alpha_{ox}$ indices and other data. Our
study shows that SBS NLS1s are strong or moderately strong soft
X-ray sources with $\, \log \,L_{x}= 42.8-45.4$. Soft X-ray
luminous sources in our sample do not tend to be luminous in the
infrared. All SBS NLS1s are radio quiet objects ($\, \log
\,L_{R}<40.0$), and 52\% of them are strong FeII emitters
(FeII$\lambda$4570/H$\beta >1$). The traditional linear
correlation $L_x$ and $L_{op}$, which seems to hold for AGN in
ge

neral, is found for SBS NLS1s. An anticorrelation between FWHM of
${H\beta}$ and the ratio of FeII$\lambda$4570/H$\beta$ is also
observed. A weak correlation is found between $\alpha_{ox}$ slope
and $L_{op}$. One of our main findings is that almost all SBS
NLS1s may not have a FIR bump. Their SED suggest that they may
also possess a BBB. The absence of IR bump in most of SBS NLS1s
and the weakness of X-ray radiation in some of them, may argue
against the presence of a BLR. The surface density of SBS NLS1s is
$<$ 0.015 per $deg.^2$ ($B<17.5, z<0.16$).

\end{abstract}


\section{Introduction}

In 1985 Narrow Line Seyfert 1 galaxies (NLS1s) were a relatively
rare and peculiar subclass of Seyfert (Sy) galaxies. NLS1s are
defined as those AGN that have an FWHM(H$_\beta$) $<$ 2000~ km ~
s$^{-1}$ and [OIII]/H$\beta$ flux ratio less than 3.0. The formal
spectral classification criteria for NLS1s galaxies is based on
(a) the presence of narrow permitted lines only slightly broader
than the forbidden ones. (b) the ratio [OIII]/H$\beta < 3$.
However, exceptions are allowed if there are also strong [FeVII]
and [FeX] emission lines, unlike to what is seen in Seyfert 2
galaxies. (c) $FWHM(H_{\beta}) < 2000 ~km~ s^{-1}$. The first two
criteria are from Osterbrock \& Pogge (1985) original
classification, while the maximum line-width criterion was
introduced by Goodrich (1989).

The properties of NLS1s are discussed by a number of authors:
Osterbrock (1977), Osterbrock \& Pogge (1987), de Grijp et al.
(1992), Bade et al. (1995), Moran et al. (1996), Greiner et al.
(1996), Wisotzki \& Bade (1997), Grupe et al. (1998, 1999), Xu et
al. (1999), Pietsch et al. (2000), etc). The 10-th edition of the
Veron-Cetty \& Veron catalogue (2001) contains data of about 150
NLS1s. Around 300 objects with published data formally might be
called NLS1s. A group of 83 NLS1s was compiled from the literature
by Veron-Cetty et al. (2001). They revised the bright part
($B<17.0$, $z<0.1$) of the NLS1s sample. High quality spectra were
obtained and a relatively large sample of 59 NLS1s was compiled.
Many objects which we would consider today to be NLS1s were not
classified as such in previous works, also some objects were
misclassified as NLS1s. This is largely due to the fact that
little effort has been made in providing a set of high-quality and
high-resolution optical spectroscopic data. Spectra do not have
enough resolution to separate unambiguously the broad and narrow
components of the Balmer lines. In addition, the presence of
strong FeII lines makes it difficult to measure H$\beta$. Previous
observations did not cover the entire wavelength range needed to
study the optical emission lines, therefore their fluxes are still
unknown. For a correct classification of the objects with complex
broad Balmer emission lines, it is also important to decompose the
H$\alpha$ and H$\beta$ emission lines into their narrow and broad
components in order to study the narrow-line (NLR) and the
broad-line region (BLR) emission ratios in NLS1s and broad line
Seyfert 1 galaxies (BLSy1s, Veron et al. 2001). Emission line
ratios from the NLR of NLS1s are different from those observed in
Sy galaxies, i.e. [OIII]/H$\beta =0.8-8$, and [OIII]/H$\beta
\geq10$ are typical ratios for the NLR of Sy2 galaxies.

The first sample that showed that NLS1s appear enhanced in X-ray
selected samples was the EINSTEIN AGN sample of Stephens (1989).
With the ROSAT All-Sky Survey (Voges 1999) it became possible to
study large samples of AGN. The ROSAT/IRAS sample of 222 AGN was
investigated by Boller et al. (1992). Bade et al. (1995) presented
the data for 283 AGN from RASS. The samples are dominated by X-ray
bright (count rate $>0.1$~ cts~ $s^{-1}$) AGN. Soft X-ray
properties for a sample of 31 NLS1s were investigated by Boller et
al. (1996). Soft X-ray properties of AGN from optical and infrared
samples were investigated by Rush et al. (1996). Moran et al.
(1996) investigated the properties of the IRAS/ROSAT sample of
AGN. Warm IRAS properties of 221 AGN were investigated by de Grijp
et al. (1992) and Keel et al. (1994). Soft X-ray properties of 76
ROSAT AGN were investigated by Grupe et al. (1998,1999). The UV
properties of the NLS1 I Zw1 were investigated by Laor et al.
(1997). ASCA observations of two NLS1s were presented by Gallagher
et al. (2001). Radio observations of a few NLS1s were presented by
Ulvestad et al. (1995), Siebert et al. (1999) and Moran (2000).
ISO (Infrared Space Observatory) observations ($7-200 \mu$) of
four NLS1s are presented by Poletta \& Courvoisier (2000).

Recent studies of QSOs with ROSAT, suggest the existence of a
significant population of soft X-ray weak QSOs where the soft
X-ray flux is $\sim10-30$ times smaller than in typical QSOs
(Brandt et al. 2000).  X-ray weak AGN were first detected with
Einstein (see Elvis \& Fabiano 1984). ROSAT extended these studies
and identified a significant population of AGN notably faint in
the soft X-ray relative to their optical fluxes
($\alpha_{ox}<-2.0$).

The proportion of NLS1s among Sy1s is still a subject of debate.
It varied from 4\% to 50\% in different samples of AGN (de Grijp
et al. 1992). A proportion of $\sim$10\% was obtained from a
medium hard (0.8-3.5 keV) X-ray selected sample of 65 AGN by
Stephens (1989). About half of the AGN in the soft X-ray selected
samples of Grupe et al. (1999) and Edelson et al. (1999) are
NLS1s. A proportion of 20-30\%, irrespective of the selection
method, was found by Engels \& Keil (2000). This last value is
often used to estimate the opening angle of NLS1s in the framework
of the unified models. For example, Taniguchi et al. (1999) prefer
to use the Stephens value, which yields an opening angle of
$\sim$10$^\circ$, because soft X-ray selected samples tend to miss
obscured AGN, but hard X-ray, radio or optical surveys are more
useful for statistical analysis.

There are no distinct physical boundaries between NLS1s and BLSy1.
Since 1985, several questions related to the NLS1s have been posed
and remain unanswered: Are they the continuation of Sy1 class
towards extreme properties? Are they accreting mass at nearly the
Eddington accretion rate? Do they have smaller black hole masses
compared to normal Sy galaxies? Are they Sy1 galaxies in their
early stages of evolution? We are still far from having an
accurate description of the main properties of NLS1s (Bade et al.
1995, Moran et al. 1996, Grupe et al. 1998, 1999, Xu et al. 1999,
Veron-Cetty et al. 2001).

Studies of NLS1s based on optically bright, strong X-ray and IR
luminous selected samples, show that they are objects with
(1)``narrow'' $<2000~ km~ s^{-1}$ broad emission lines, (2) strong
optical FeII and weak [OIII] lines, (3) a very steep soft X-ray
spectra (Laor et al. 1994,1997, Boller et al. 1996) together with
higher soft X-ray luminosities and (4) strong IR emitters (Halpern
and Oke 1987; Moran et al. 1996; Lipari 1994).  NLS1s are found to
be AGN with extreme properties: they show the steepest X-ray
spectra, the strongest FeII emission and the lowest emission from
the NLR. Furthermore, they are more variable in X-ray and have
stronger FeII emission than BLS1s. NLS1s are found to be radio
quiet objects with compact radio sources in their central
($<300$~pc) zone. Mid and far-infrared properties were found
similar to those of BLS1s. A summary of selected important
observational properties of NLS1s in the optical, UV, soft and
hard X-ray, radio and infrared, was presented by Taniguchi et al.
(1999). Recent data on weak soft X-ray sources and radio-loud
NLS1s, and new data\footnote{Recently, Rick Williams, Dick Pogge
and Smita Mathur published a sample of 150 SLOAN NLS1s
(astro-ph/0208211). This sample confirms the presence of NLS1s
which are not X-ray sources, which is part of the results founded
within our SBS NLS1s sample. Obviously, the SLOAN NLS1s sample
expanded the investigation of NLS1s to a fainter limit. We will
analyze the SLOAN NLS1s sample and also other NLS1s samples and
present our results in a separate paper.} may significantly change
their characterization.

Weak X-ray NLS1s have not yet been studied in detail, nor
homogeneously. A few dozen of them were classified in optical
samples of AGN, a dozen were found among radio and infrared
sources, and a relatively large fraction of NLS1s was found among
the soft X-ray ROSAT sources. Therefore, the question is if the
obtained results are a consequence of using  inadequate samples.
We will show below, that some results are indeed sample dependent.

Multiwavelength investigations of NLS1s are relatively poor due to
the difficulty in building a representative multiwavelength
complete sample of NLS1s. Building this sample  requires a
representative multiwavelength complete sample of AGN. AGN are
predominantly selected in optical (single waveband) surveys,
commonly due to their UV excess (UVX, First Byurakan Survey,
Markarian 1967; Second Byurakan Survey [SBS], Markarian \&
Stepanian 1983); Bright Quasar Survey, Schmidt \& Green 1983;
Large Bright Quasar Survey, Hewett, Foltz \& Chafee 1995; etc). In
general optical surveys do not miss any UVX AGN. The radio surveys
of AGN are largely free from color bias. Hard X-ray emission is
relatively unaffected by absorption, while the soft X-ray surveys
are strongly affected by it. Multiwavelength observation of AGN
over the past decade led to fundamental improvements in our
understanding of the energy generation mechanisms in these
objects. AGN are multi-wavelength emitters, with roughly equal
energy output extending from the FIR through to the X-ray region.
Single waveband studies cannot discriminate the general properties
of the parent population of NLS1s (or other types of AGN). Of
course the combination of the X-ray, optical, FIR, radio and other
waveband surveys in the same sky area may produce a more complete
sample of AGN.   Such a sample in principle may answer the
question of the proportion oNLS1s among the X-ray, optical, IRAS
and radio selected AGN.

During the Second Byurakan survey a sample of 578 QSOs and AGN was
spectroscopically confirmed (Stepanian 1994; Stepanian \&
Chavushyan 2003). In this paper we present a multiwavelength study
of the NLS1s sample (26 objects) extracted from the SBS survey,
which is homogeneous in a limited sky area.

\section{The SBS NLS1s}

To isolate the NLS1s sample from the SBS general sample of AGN
(Stepanian 1994; Stepanian \& Chavushyan 2003) we formally follow
the two optical spectroscopic criteria described above. Among the
578 AGN from the SBS, $\sim50$ showed ``narrow'' emission lines
and emission line properties which satisfy the NLS1s criteria. The
magnitudes, redshifts and luminosity intervals of these objects
correspond to the following values: $15.2< B < 19.0$, $0.02
<z_{em} <0.65$, $-19.9 > M_{B} > -26.5$.

To be consistent with the optical definition for QSOs and Sy
galaxies (Schmidt \& Green 1983), we formally divided our original
sample into two sub-samples in terms of their optical luminosity.
The first one with $M_{B}\ge -23.0$ ($-19.9 < M_{B}
>-23.0$) is called the NLS1s sample. The second one with $M_{B} <
-23.0$ ($-23.0 > M_{B} > -26.5$), is the narrow-line QSOs sample
(NLQSOs).

In this paper, we will present only the study of the NLS1s sample
from the SBS (hereafter SBS NLS1s). The multiwavelength data and
the corresponding study of  SBS NLQSOs will be discussed in a
forthcoming paper.

\subsection{The sample of SBS NLS1s galaxies}

The SBS catalogue contains 26 NLS1s galaxies. In Table \ref{t1} we
present the basic data for this sample as follows: Col. (1) SBS
designation (equinox B1950), according to the IAU nomenclature;
columns (2) and (3) J2000.0 coordinates from Bicay et al. (2000),
their accuracy is $\pm$1\arcsec; column (4) 1RXSJ designation
according to Voges et al. (1999); columns (5) and (7) distance in
arcsec of the ROSAT and radio sources from the SBS optical
objects, for some objects according to Schwope et al.(2000), Bauer
et al. (2000) and Bade et al.(1995); column (6) FIRST (Faint
Images of the Radio Sky at Twenty cm) or NVSS (NRAO VLA Sky
Survey) designation according to Becker et al. (1995) and Bauer at
al. (2000); column (8) near infrared data, according to the Nasa
Extragalactic Database [NED]; column (9) the optical major and
minor axis in arcsec; columns (10) and (11) rough estimates of
FeII strength complexes, $\lambda$4570 and $\lambda$5100,
classified as w (weak), s (strong), and vs (very strong); column
(12) other name; and column (13) references.

We find useful to present in Table \ref{t1} the estimation of the
strength of FeII complexes $\lambda$4570 and $\lambda$5100. The
emission lines of FeII consist of hundreds of energy levels in
many multiplets, which make them very complex. Boroson \& Green
(1992) and Bade et al. (1995) have tried to quantify the strength
of the FeII emission by measuring the equivalent widths of two
FeII blends between $\lambda$4400-4680 \AA~ (Boroson \& Green used
the wavebands $\lambda$4434-4684 \AA~ ) and $\lambda$5100-5500
\AA. These estimates introduce errors because HeII$\lambda$4686
\AA~ is also present in this blend. Nevertheless, we will use this
estimate to quantify the strength of the FeII emission in our
sample.

\subsection{Identification}

In this section we discuss the cross identification of optical SBS
sources with sources at different wavelengths. To identify an
object as a candidate source for ROSAT, IRAS or radio we have used
the distance between optical position from the SBS and the source
positions found in the other surveys.  Each SBS object has an
accurate measured coordinate ($\leq 1$\arcsec) and we defined
these coordinates as the central position. Then, the X-ray, IRAS
and radio positions were used to calculate the distance of these
sources from the optical position.

All objects were identified with 1RXSJ sources (Voges et al. 1999,
2000), except four: SBS 0924+495, SBS1136+595, SBS 1509+522 and
Mkn 486(1535+547), while Mkn 124, SBS 0952+522, SBS 1359+536, and
SBS 1406+540 were identified as RASS faint sources (Voges et al.
2000). Many of them are identified here as 1RXSJ sources for the
first time. Objects identified as X-ray sources have their
position coincident with the optical SBS objects, i.e. within
$\leq 19$\arcsec.

Only one object, Mkn 110 (0921+525), was identified as an IRAS
source (F09213+5232). The difference in distance between the
optical and IRAS position is 19\arcsec.

Nine objects were identified as radio sources, eight of them are
FIRST sources. All FIRST sources have a position coincident by
less than 1\arcsec. SBS 1315+604 was identified as NVSS source by
Bauer et al. (2000) with a position difference of 16\arcsec.

The galaxy SBS 1213+549A was misidentified by Boller et al.
(1992), Moran et al. (1996) and by Condon et al. (1998) as
MCG09-20-133, and as an IRAS source. The same mistake is included
in other databases and the catalogue of Veron-Cetty \& Veron
(2001). This misidentification is understandable because there are
three different galaxies in this field: SBS 1213+549A, MCG
09-20-133 and SBS 1213+549B (MCG 09-20-134).

A cross-identification is presented in Table \ref{t2}. The IRAS,
ROSAT, optical, and radio positions for the B1950 and J2000
epochs, according to Saunders et al. (2000), Voges et al. (1999),
Becker et al. (1995), Condon et al. (1998a, 1998b), Moran et al.
(1996) and Boller et al. (1992) are presented for SBS 1213+549 A
and MCG 09-20-133. The difference in coordinates between the ROSAT
(1RXSJ121549.3+544227) and the optical SBS 1213+549A source is
3\arcsec; between the optical and radio source
(FIRSTJ121549.4+544223) is less than $1\arcsec$. The difference
between the IRAS and SBS 1213+549 A coordinates is 42\arcsec. The
difference in coordinates of IRAS and the other galaxy, in the
same field of MCG 09-12-133 is 17\arcsec. Therefore, SBS 1213+549A
is more likely to be a ROSAT and FIRST source, which was certainly
not detected by IRAS. We think that MCG 09-20-133 might be the
IRAS 12134+5459 source, and that it probably is not a ROSAT nor a
FIRST source, but is more likely to be a NVSS source.

In Bade et al. (1995), 68\% of the RASS/IRAS sources were found
within 20\arcsec. Moran et al. (1996) find that 70\% of the
optical counterparts of RASS sources lie within 20\arcsec of the
RASS X-ray position. In our sample, all SBS NLS1s have a distance
coincidence with RASS sources less than 19\arcsec. This gives us a
reliable identification of the objects, which will be useful for
further statistical investigation of their multiband properties.

\section{Spectra and Classification}

\subsection{Classification}

The objects presented in Table \ref{t1} were originally classified
as Sy1, Sy1.5, Sy1.8, Sy1.9 or Sy2 in previous works, e.g. see
NED. Five of them (Mkn 110, Mkn 124, Mkn 142, Mkn 486, SBS
1213+549A) were classified as NLS1s by Osterbrock \& Pogge (1987),
Martel \& Osterbrock (1994),  Veron et al. (2001) and Moran et al.
(1996). Our spectroscopic observations of 26 objects presented
below have shown that according to the $FWHM(H_{\beta})<2000$~km~
$s^{-1}$ and [OIII]/H$_{\beta}<3.0$ criteria, these objects are
NLS1s.

\subsection{Observations}

The spectroscopic observations of the SBS sample were carried out
with the 6 m telescope of the Special Astrophysical Observatory
(SAO, Russia). The SP-124 spectrograph, equipped with a
1024-channel image photon counting system (IPCS) (Drabek et al.
1986)was used, installed in the Nasmyth focus of the telescope,
and also a spectrograph equipped with a $580 \times 530$ pixel CCD
(Afanas'ev et al. 1995), installed at the prime focus. The
instrumental spectral resolution is $\sim 6$ \AA, in the
wavelength interval 3500-7200 \AA. Another set of data was
collected with 2.1 m telescope of the Guillermo Haro Observatory
(GHO) located at Cananea, Sonora (M\'exico). We used the LFOSC
spectrophotometer installed in the Cassegrain focus (Zickcraf et
al. 1997), equipped with a $578\times385$ pixel CCD, which gave us
an effective instrumental spectral resolution of $\sim 11$ \AA, in
the wavelength interval 4000-9000 \AA. A new set of spectroscopic
observations were carried out both at GHO and also at the
Observatorio Astron\'omico Nacional in San Pedro M\'artir
(M\'exico) with the 2.1 m telescope. A high signal-to-noise ratio
($S/N\sim30-50$) and better spectral resolution of ($\sim3.5-4$
\AA) in the waveband 3700-8300 \AA, allow us to confirm the
previous spectral classification.

The emission line parameters (EW, FWHM, emission line ratio) were
determined using the spectral analysis software package developed
by V. V. Vlasyuk (1995, private communication) at the Special
Astrophysical Observatory. The fluxes and equivalent widths of two
FeII blends were measured between $\lambda$4400-4680 \AA  and
$\lambda$5100-5500 \AA. The data are not corrected for reddening.
For our measurement the instrumental broadening corrected value is
given (i.e. by subtraction of the night sky profiles).

The results of the spectroscopic observations of SBS NLS1s are
presented in Table \ref{t3}. Data collected from the literature is
presented as well. Columns are as follows: (1) SBS designation;
(2) to (9) EW and the FWHM of H$\alpha$, H$\beta$, H$\gamma$, and
[OIII]$\lambda$5007, respectively; (10) and (11) EW of
FeII$\lambda$4570=F(4434-4684) or F(4500-4680) if the data are
taken from Bade et al. (1995) and FeII$\lambda$5100=F(5100-5500);
(12) range of the FWHM of H$\beta$ according to Veron et al.
(2001); (13) references. The observed EW as well as FWHM in Table
\ref{t3} have been corrected to z=0, i.e. $W(\lambda,z)\,=\,
W(\lambda,z=0)\, (1+z)$.

The corresponding data of the emission line ratio are summarized
in Table \ref{t4}; the columns are as follows: (1) is the SBS
designation; (2) [OIII]$\lambda$5007/H$\beta$; (3)
H$\alpha$/H$\beta$; (4) FeII$\lambda$4570/[OIII]; (5)
$R(4570)=FeII\lambda4570/H\beta$ and (6) references. The
uncertainty in the intensity ratios for strong emission lines
reported here is about 25\% (for our observations), and
$\sim25-30\%$ for data obtained in the literature.

According to the data presented in Tables \ref{t3} and \ref{t4},
we may assume that the SBS NLS1s sample satisfies the formal
criteria for NLS1s galaxies. Only Mkn 110 (SBS 0921+525) has the
FWHM of H$\beta$ between 1670-2500, with the median value greater
than 2000 ~km ~$s^{-1}$.

\section {The multiwavelength data}

Since we believe that in order to deal properly with
multiwavelength data, authors should present their definitions,
conversion factors and calculations, we include them in this
section. We have adopted: $q_{0}=0$, $H_{0}=50$ $km~ s^{-1}$~
$Mpc^{-1}$.  ROSAT fluxes $f_{x}$ are in units of $10^{-12}$~ erg~
$cm^{-2}$~ $s^{-1}$, IRAS fluxes $F_{FIR}$ are in Jy, radio fluxes
S(R), in mJy. The luminosity for all bands $L_{opt}$, $L_{FIR}$,
$L_{x}$, $L_{R}$ is given in common units of $erg~s^{-1}$.

The luminosity in each band was calculated according to the
expression: $L_{\lambda}\, = \,4 \pi \, D_{z}^2 \,  F_{\lambda}$,
where $F_{\lambda}$ are the fluxes in optical, X-ray, radio and
FIR bands and $D_{z}$ is the distance of the object given by the
expression:

\begin{equation}
D_{z}= \frac{c}{H_{0}} A(z)\,=\,\frac{c}{H_{0}}\,\frac{q_{0} z
+(q_{0}-1)[(1+2q_{0})^1/2 - 1]}{q_{0}^2}.
\end{equation}
For $q_{0}=0$, $A(z)\,=\,z(1+z/2)$ and $ L_{\lambda}\,
=\,4\pi(c/H_0)^2  [z(1+z/2)]^2  F_{\lambda}$.

Table \ref{t5} contains the multiwavelength data in X-ray, optical
and radio regions for SBS NLS1s sample: column (1) SBS
designation; column (2) redshift, derived from the mean value of
the redshifts of the strongest emission lines; column (3) B
magnitudes with two decimals have an accuracy of $\pm0.05^{mag}$,
while the others have $\pm 0.5^{mag}$; column (4) absolute
magnitude $M_{B}$, defined by $M_{B}$ = B - 5$\, \log \,[z(1+z/2]$
+ $2.5(1+\alpha)\, \log \,(1+z)$ - 43.89, we used $\alpha$ = -0.7
for the K correction, which is negligible for low redshift
objects; column (5) the $\, \log \,L_{B}$= 53.03 - 0.4 B + 2$\,
\log \,[z(1+z/2)]$; column (6) X-ray counts according to Voges et
al. (1999); column (7) is the X-ray flux density according to
Voges et al. (1999; 2000), the units are $10^{-12} erg~ s^{-1}~
cm^{-2}$; column (8) $\, \log \,$ of the monochromatic flux
densities at 2 keV in $erg~ cm^{-2}~ s^{-1}~Hz^{-1}$; column (9)
$\, \log \,$ of the soft X-ray luminosity, $\, \log \,L_{x}$ =
57.63 + $2\, \log \,[z(1+z/2)]$ + $\, \log \,f_{x}$; column (10)
the radio flux density at 1.4 GHz (20 cm), the peak intensity
value in mJy according to Becker et al. (1995) and Bauer et al.
(2000); column (11) $\log \,L_{R}$  at 1.4 GHz, given by $\log
\,L_{R}$ = 40.78 + $2\, \log \,[z(1+z/2)] \,+ \, \log \,S(mJy)$;
column (12) $\, \log \,(f_{x}/f_{opt})$ = $\, \log
\,[f(2keV)/f(\lambda4000$ \AA)]= 0.4B + 1.52 + $\log \,f_{x}$,
calculated for the ROSAT soft energy band 0.05 -- 2.4 keV., and
$\alpha$ = -0.5, according to the equation [1] of Schmidt \& Green
(1986); and column (13) $\alpha_{ox}$ = 0.372~$\, \log
\,[f(2keV)/f(3000$~\AA)] the slope of a power law defined by the
rest-frame flux densities at 3000 \AA~ and 2 keV (Brandt et al.
2000).

Our blue magnitudes were converted to monochromatic flux densities
at 3000 \AA~ and are calculated according to Schmidt \& Green
(1986):

\begin{equation}
   B\, =\,- 2.5 \, \log \,f(\lambda) + 1.25 \, \log \, \frac{\lambda}{4400}~ -
   48.36.
\end{equation}
In all cases, B is the magnitude at 4400 \AA.

Monochromatic flux densities at 2 keV, $f_{x}(2keV)$ are derived
from the fluxes $f_{x}$(0.05, 2.4 keV) according to Schmidt \&
Green (1986);

\begin{equation}
f_{x}(2keV)=4.14 \times 10^{-18}(1+\alpha)
2^{\alpha}\frac{F_{x}(0.05,2.4keV)}{2.4^{1+\alpha}-
0.05^{1+\alpha}}
\end{equation}

The monochromatic fluxes in the optical band at $\lambda$4400 \AA,
are calculated according to Allen (1976);

\begin{equation}
f_{opt}\,=\, 10^{-0.4\,B \,-\, 19.34}.
\end{equation}

The units in equations (3) and (4) are erg~$cm^{-2}$~ $s^{-1}$~
$Hz^{-1}$.

For 10 objects of Table \ref{t5}, the multivawelength data in
different wavebands are presented also in de Grijp et al. (1992),
Bade et al. (1995), Boller et al. (1996), Moran et al. (1996),
Rush et al. (1996) and Brandt et al. (2000). In all of them,
optical and X-ray fluxes and luminosities, and $\alpha_{ox}$
indices differ from each other, because of the difference in
magnitudes, redshifts, count rates and other parameters used. We
have used in Table \ref{t5} our original measurement of redshifts
and B magnitudes. The soft X-ray count rates, ROSAT and radio
fluxes were taken from the last version of original catalogues,
Becker et al. (1995) and Voges et al. (1999). For four objects, we
used data from the ROSAT faint source catalogue (Voges et al.
2000), marked with an asterisk.

Note that in Table \ref{t5}, the absolute magnitude $M_{B}=-23.0$
was used to isolate the SBS NLS1s sample, this value corresponds
to the optical luminosity $L(B)\,=\,$44.70~ erg~ s$^{-1}$, which
is our formal division limit to discriminate between NLS1s and
NLQSOs by luminosity.

In Table \ref{t6} we show the following data for Mkn 124,
IRASF09213+5232 source, with column (1) the SBS designation;
columns (2), (4), (6), (8) the IRAS fluxes at 12, 25, 60 and 100
$\mu$m in Jy; columns (3), (5), (7), (9) the quality of detection,
where 1- upper limit, 2 - medium, 3 - high quality detection;
columns (10), (11), (12) the FIR colors, i.e. the logarithms of
the FIR flux ratios; column (13) $\log \,F_{FIR}$, where $F_{FIR}$
= 1.26 $\times 10^{-14} [2.58 F(60Jy) + F(100Jy)]$ with FIR fluxes
in $W~ m^{-2}$, according to Bicay et al. (1995); column (14) $\,
\log \,[L_{FIR}/L_\odot]$ = 27.051 + 2\, $\log$ \,[z(1+z/2)] + \,
$\log$ \,$F_{FIR}$; and column (15)  $\log$ \,$L_{FIR}$.

\begin{figure}

\figurenum{1} \plotone{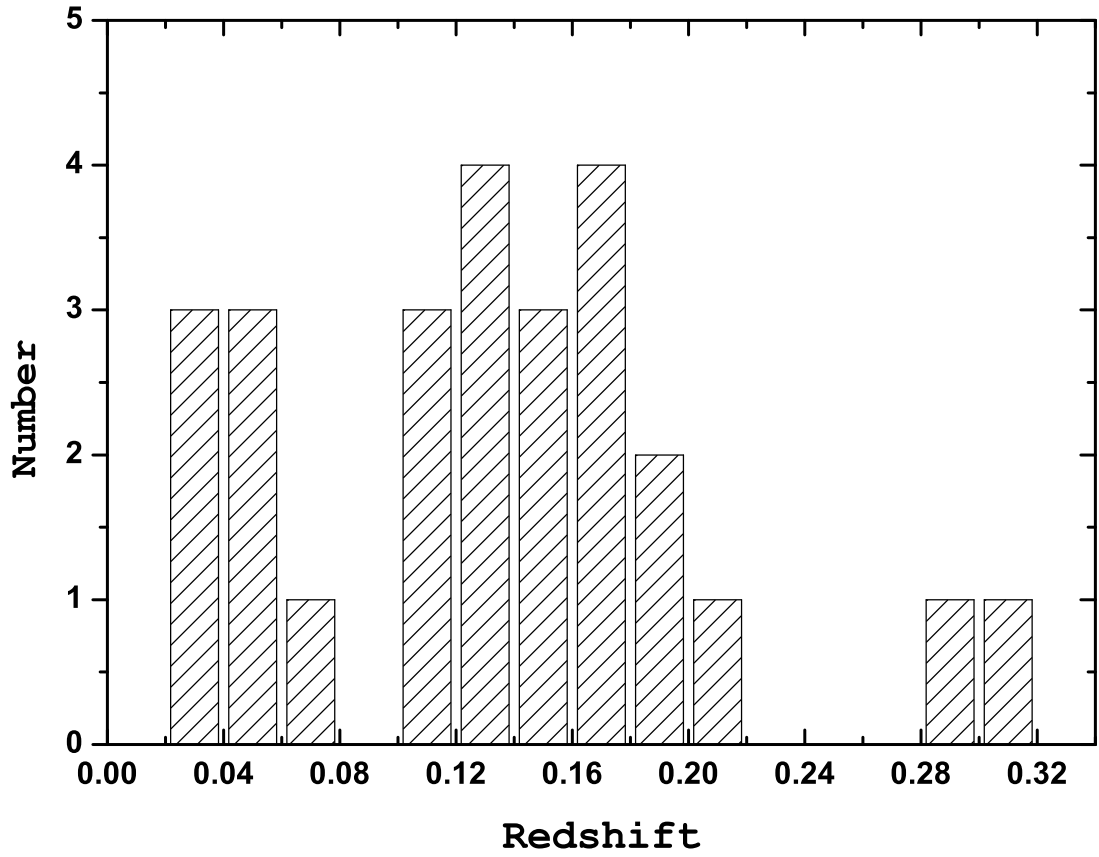} \caption[f1.eps]{Redshift
distribution of SBS NLS1s sample. \label{fig1}}

\end{figure}

\begin{figure}

\figurenum{2a} \plotone{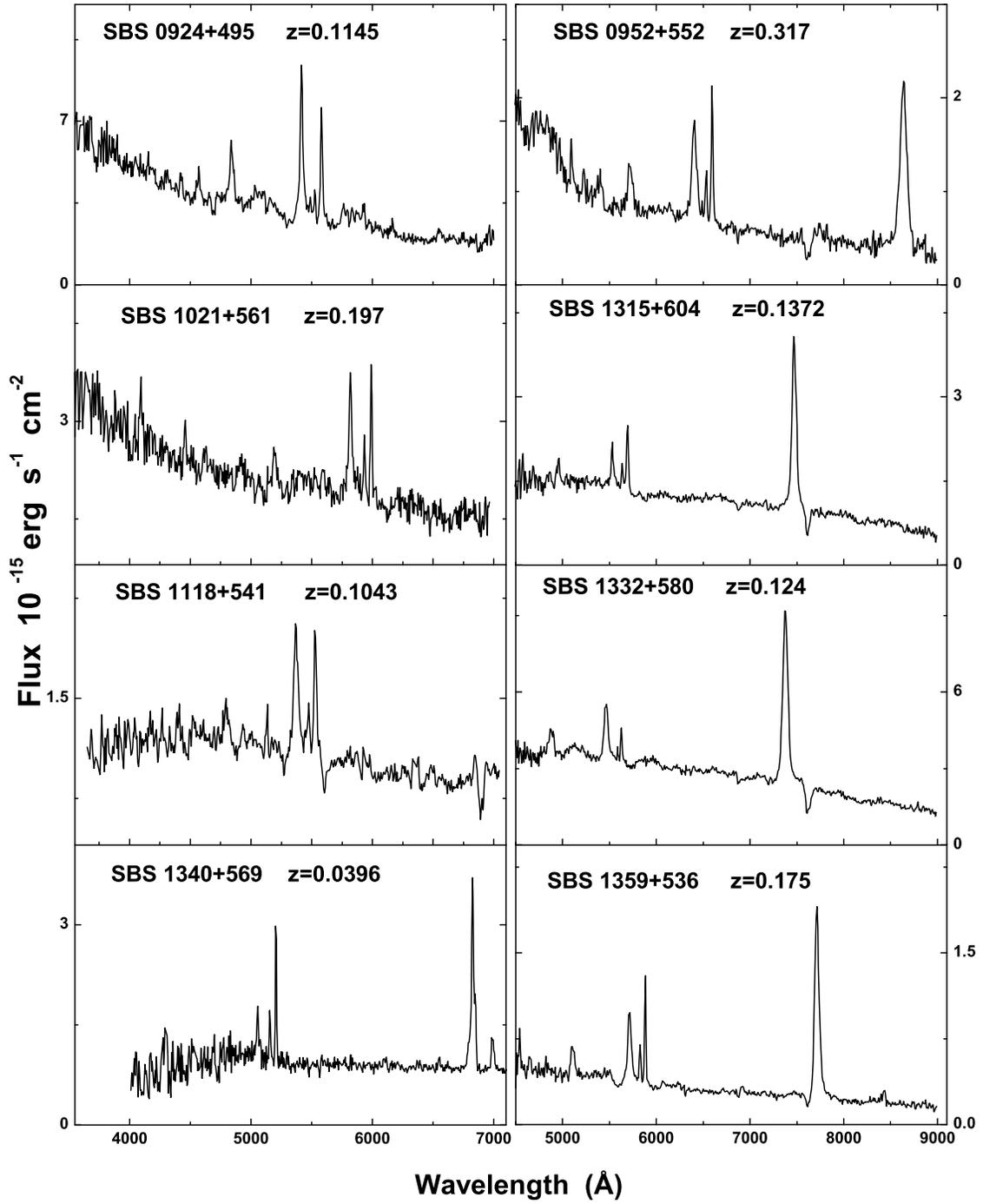} \caption[f2a.eps]{Spectra of SBS
NLS1s. Only objects without a former published spectra are
presented. \label{fig2a}}

\end{figure}

\begin{figure}

\figurenum{2b} \plotone{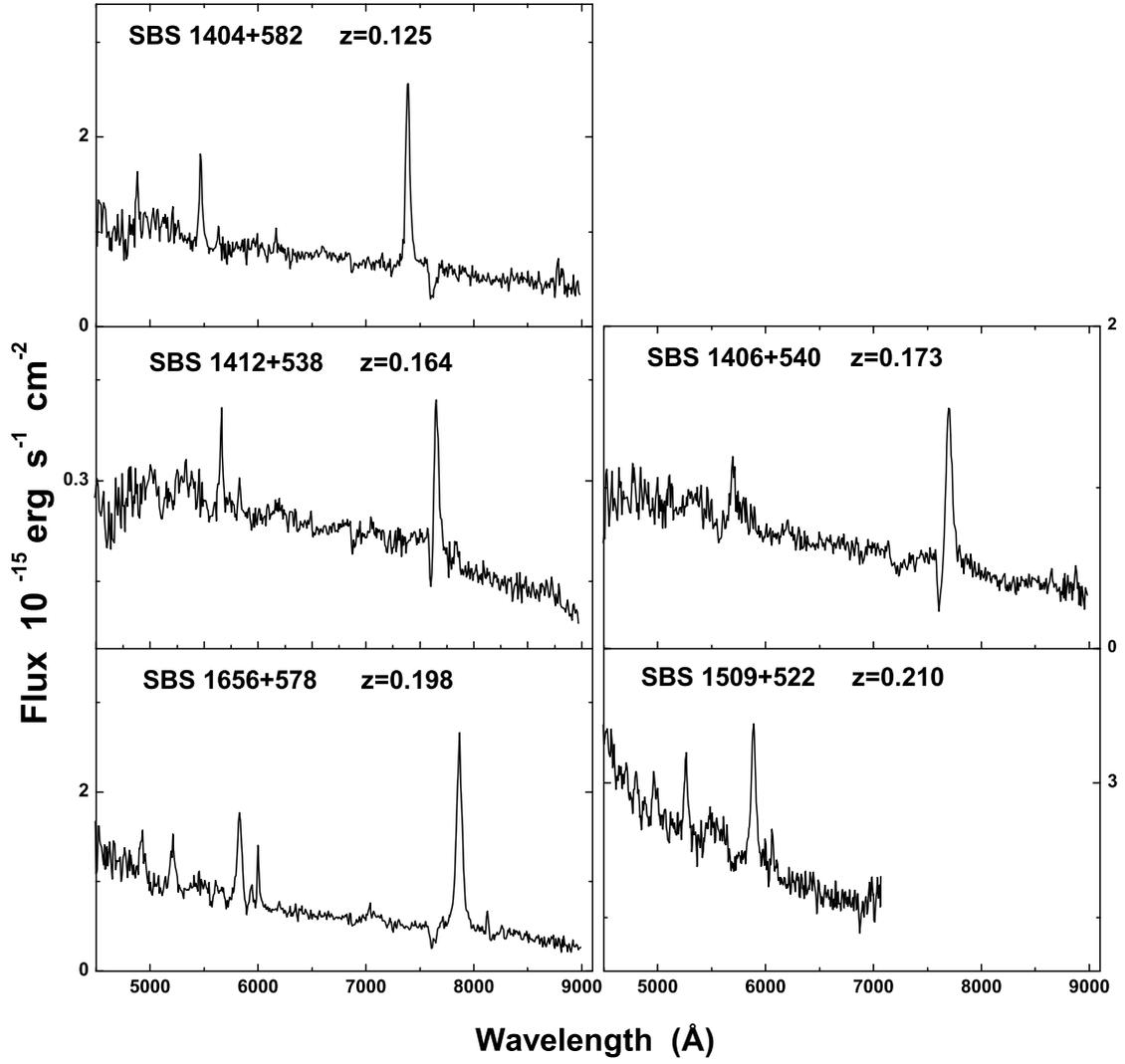} \caption[f2b.eps]{Spectra of SBS
NLS1s. Continued. \label{fig2b}}

\end{figure}

The redshift distribution of SBS NLS1s is shown in
Figure\ref{fig1}. The optical spectra of 13  SBS NLS1s is given in
Figures \ref{fig2a} and \ref{fig2b}, since these objects lack
previous spectroscopic observations. The optical spectra of Mkn
110, Mkn 124, Mkn 142, Mkn 486, SBS 1213+549A and some others,
obtained in different wavebands with different spectral
resolution, might be found in Osterbrock (1977), de Robertis
(1985), Stephens et al. (1989), Boroson \& Green (1992), de Grijp
et al. (1992), Martel \& Osterbrock (1994), Bade et al. (1995),
Moran et al. (1996), Grupe et al. (1999) and Veron et al. (2001).
We present the relevant properties of the SBS NLS1s galaxies in
section 5.

\section{Results}

\subsection{Continuum and emission line correlations}

Principal Component Analysis (PCA) is a valuable tool to study the
relationship between several variables. Unfortunately, PCA is not
reliable for the SBS NLS1s sample presented in this paper, because
of the small number of objects. Furthermore, the missing and upper
limits values in the sample reduce even more the available number
of objects when several important variables are taken into account
for the analysis. Thus, we have compared 13 of the variables
included in Tables 3, 4 and 5, in groups of two yielding a total
of 78 comparisons. For such a large group of comparisons, we
expect to obtain around 4 spurious correlations at the 0.05 level
of significance. Table \ref{t7} shows the relevant correlations at
this level. The 13 correlations listed involve 12 out of the 13
variables considered. Column (1) in Table \ref{t7} assigns a
number for each comparison of variables; column (2) lists the
variables compared; column (3) indicates the number of objects
involved in the comparison; column (4) shows the probability that
the correlation or anticorrelation obtained has been drawn by
chance from unrelated variables; and column (5) indicates the
correlation coefficient for each pair of variables.

\begin{figure}

\figurenum{3} \plotone{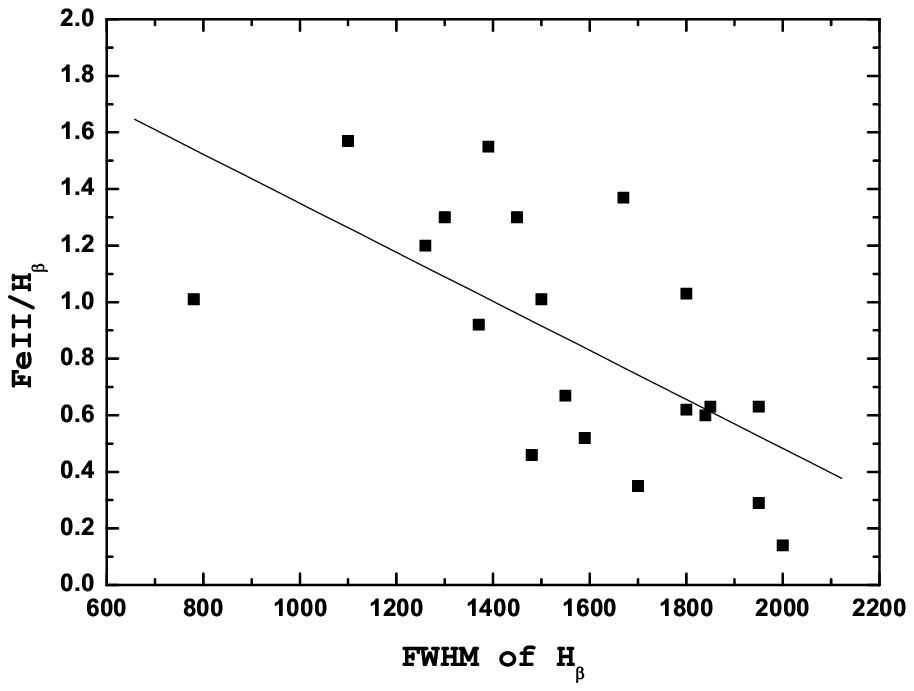} \caption[f3.eps]{The FWHM of
H$\beta$ versus FeII$\lambda$4570/H$\beta$ ratio for SBS NLS1s.
The solid line stands for the best possible fit to the points.
\label{fig3}}

\end{figure}

A few patterns may be drawn from Table \ref{t7}. Comparisons 1 to
4 and 7 involve either EW of H$\beta$ or FWHM of H$\beta$, or
both. Variables EW of H$\beta$ and FWHM of H$\beta$ seem to
measure the H$\beta$ power (Boroson \& Green 1992, eigenvector 1),
as can be inferred from the anticorrelations of both variables
with FeII/H$\beta$. Furthermore, anticorrelations 3 (presented in
Figure\ref{fig3}), 4, and 5 (and probably 2 and 7), and
correlations 6, 8, 9, and 13 (and probably 10) are related with
Boroson \& Green's eigenvector 1. However, when using radio
luminosities we are restricted to 10 objects, so these have to be
taken cautiously. This is evident for correlation 9, where the
objects with the largest [OIII]/H$\beta$ ratios (SBS 1340+569 and
SBS 1353+564) have not been radio detected.

Correlations 8 and 12 are associated with continuum emission from
the AGN, and probably indicate related mechanisms for the energy
production. Correlation 13 indicates that variables $\alpha_{ox}$
and $\, \log \,L_{x}$ measure the X-ray power. In the same way,
correlation 11 shows that variables $\, \log \,(L_{R}/L_{x})$ and
$\, \log \,L_{R}$ measure radio power.

Relations between line ratios are considered in comparisons 5 and
6. Anticorrelation 5 is produced because of the different place
(numerator and denominator) of the [OIII] flux in the line ratios
involved. A similar explanation is also valid for correlation 6,
where FeII is in the same place for both line ratios. The fact
that the ratio FeII/[OIII] is involved in comparisons 5 and 6
reflects that this line ratio is linearly dependent on the other
two line ratios: FeII/[OIII] = (FeII/H$\beta)$ /([OIII]/H$\beta)$.
A correlation between [OIII]/H$\beta$ and FeII/H$\beta$ would also
be expected for analogous reasons (H$\beta$ power dividing [OIII]
and FeII powers in both cases), but the [OIII]/H$\beta < 3$
condition imposed in the selection of the NLS1s sample hides this
correlation.

\subsection{X-ray and optical (B) continuum correlation}

\begin{figure}

\figurenum{4} \plotone{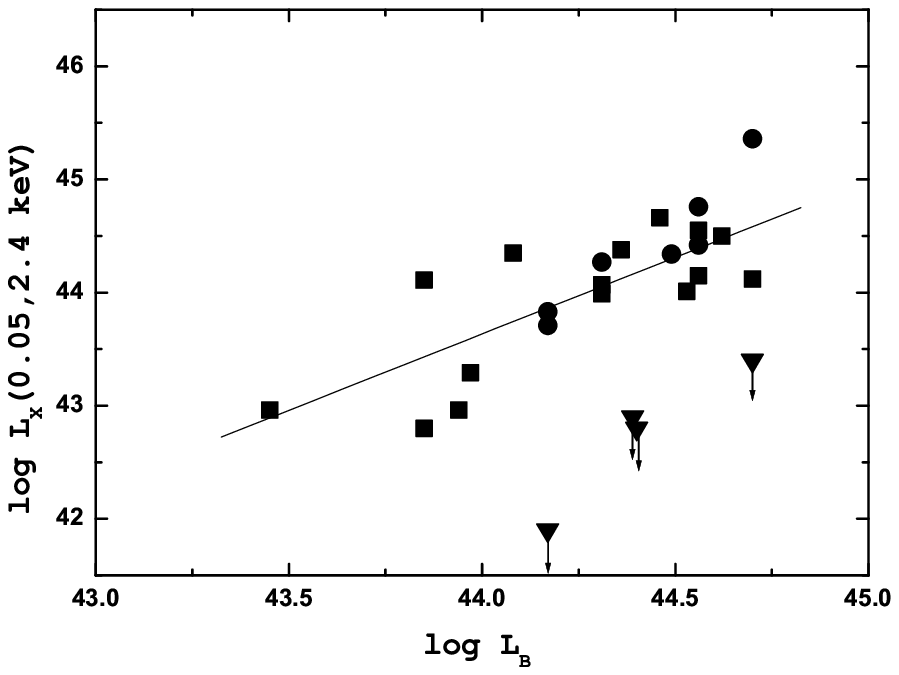} \caption[f4.eps]{The diagram of
optical versus X-ray luminosity of SBS NLS1s for all the objects
and restricted by magnitude $B\le17.5$ ($\blacksquare$). The line
represents the best fit. Weak sources ($B>17.5$,
{\LARGE$\bullet$}) are also shown. Four objects without a
detection ($\blacktriangledown$) were excluded from the analysis.
\label{fig4}}

\end{figure}

\begin{figure}

\figurenum{5} \plotone{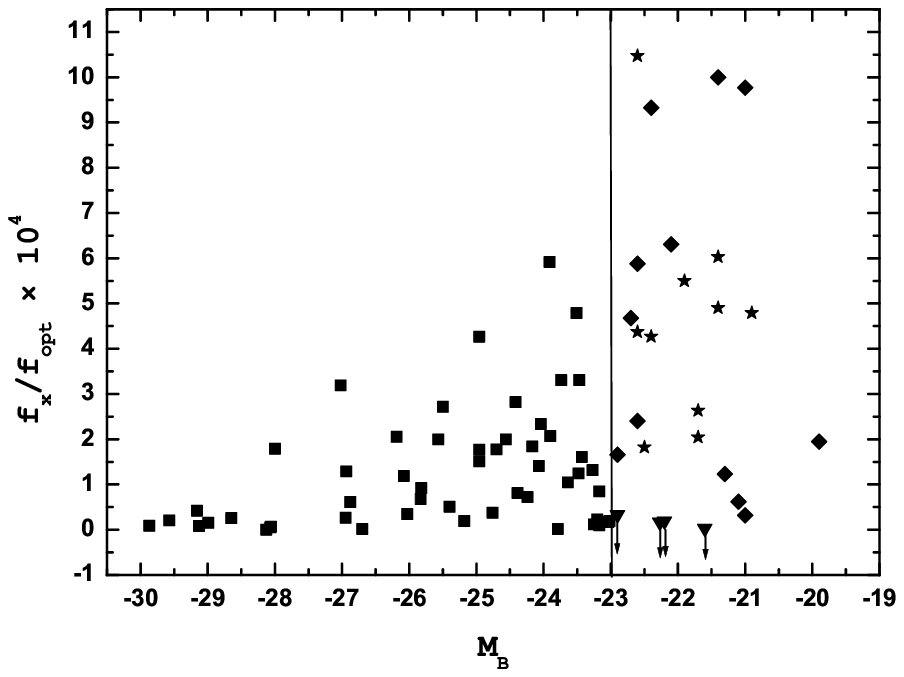} \caption[f5.eps] {The diagram of
the $f_{x}/f_{opt}$ as a function of $M_{B}$ for SBS NLS1s
($-19.0>M_{B}>-23.0$, $B<17.5$, $\blacklozenge$, N=14), added to
the same diagram of BQX QSOs of Schmidt \& Green (1986)
($\blacksquare$) ($-23.0>M_{B}>-30.0$, N=52). Fainter objects
($B>17.5$, $\bigstar$) and upper limits ($\blacktriangledown$) are
also represented.

\label{fig5}}

\end{figure}

In Figure\ref{fig4} we present the diagram of optical versus X-ray
luminosity for the total and restricted by magnitude
$B\le17.5$\footnote{The mag. limit $B\le17.5$ means that only the
members of the complete sample of SBS NLS1s were considered. }
(shaded squares) sample of SBS NLS1s. A diagram showing
$f_{x}/f_{opt}$ as a function of $M_{B}$ is shown in
Figure\ref{fig5}. In all cases, four objects not detected in soft
X-ray band were excluded and are marked be arrows in
Figure\ref{fig4} and \ref{fig5}.

A strong linear correlation between X-ray and optical luminosity
is seen for SBS NLS1s detected in X-ray. From the data plotted in
Figure \ref{fig3}, we obtained the following results after
applying a linear regression fit:

\begin{equation}
\, \log \,L_{x}\, = \,1.34(\pm0.3) \, \log \,L_{B} - 15.3(\pm3.7),
\end{equation}

with r=0.77 and  N=14.

We note that a similar correlation was found by Boller et al.
(1992), $\, \log \,L_{x}$=1.59 $\, \log \,L_{B}$ - 29.24, r= 0.72)
for 20 X-ray bright AGN from the ROSAT/IRAS survey (this sample
contains a mixture of different types of Sy galaxies).

A correlation $L_{x}$ - $L_{opt}$ seems to hold for AGN in general
(La Franca et al. 1995). It is believed that hard X-ray emission
is reprocessed into the soft X-ray-to-optical range by cold dense
clouds. The correlation between X-ray and optical luminosities
suggest that the mechanisms producing soft X-ray  and optical
emission are strongly related.

The dependence of $f_{x}/f_{opt}$ of AGN on redshift, optical and
radio luminosity has been extensively discussed in the literature.
Trends in the spectral shape have been investigated by Schmidt \&
Green (1986), for a sample of 53 Bright Quasar X-ray Sample (BQX)
QSOs($M_{B}<-23.0$). They found that $f_{x}/f_{opt}$ increases
toward fainter optical absolute magnitudes, down to $M_{B}$ =
-23.0, with increasing scatter. Boller et al. (1992), confirm the
result obtained by Schmidt \& Green for QSOs, finding a linear
correlation between $f_{x}/f_{opt}$ and $M_{B}$, with slope
$\alpha=0.30$, and correlation coefficient r=0.47.

We studied the relation between $f_{x}/f_{opt}$ and $M_{B}$ for
both the total and $B\le$17.5 sample of SBS NLS1s. The sample of
SBS NLS1s does not show the correlation between $f_{x}/f_{opt}$
and $M_{B}$ in a luminosity range -19.0$\,<\,M_{B}\,<\,$-23.0,
$\alpha$ = 0.1, r=0.07. A similar result was found by Boller et
al. (1992): $\alpha$ = 0.02, r=0.04. If the sample of SBS NLS1s
($B\le17.5$) is added to the sample of QSOs of Schmidt \& Green
(1986), then their trend, i.e. the increasing of the spread of
$f_{x}/f_{opt}$, might be continued up to the optical luminosity
of $M_{B}$ = -19.0 (see Figure\ref{fig5}).

The correlation between $f_{x}/f_{opt}$ and $M_{B}$ needs further
detailed investigation since it is used to describe the spectral
behavior of extragalactic objects. $f_{x}/f_{opt}$ vs. $M_{B}$ or
$L_{x}/L_{opt}$ statistics as a function of optical luminosity,
are useful for the modelling AGN number counts, to predict the
X-ray quasar counts and to estimate the X-ray background in the
Universe.

The mean value of $\overline{\, \log \,(f_{x}/f_{opt})}$ is
$-3.64\pm0.4$ for SBS NLS1s including the limiting value estimate,
and -3.43 only for detected sources. The mean value of
$\overline{\, \log \,(f_{x}/f_{opt})}$ for 143 AGN (119 of them
are Sy1s) determined by Boller et al.(1992) is $-4.22\pm0.63$.
Thus, this means that SBS NLS1s are either much weaker soft X-ray
sources, or optically more luminous objects than the ROSAT AGN
investigated in Boller et al. (1992).

\subsection{Spectral energy distribution (SED) and $a_{ox}$ slope}

Spectral energy distribution (SED) studies of AGN increased
rapidly in the past decade. Nevertheless, there are very limited
data in the radio, FIR, sub-mm, extreme UV and $\gamma$ ray
regions. The FIRST, NVSS and ROSAT surveys provided data in radio
and soft X-ray regions, the ISO (Infrared Space Observatory),
SIRTF (Space Infrared Telescope Facility) and SCUBA (Submm
Common-User Bolometer) should provide new information in the IR
and sub-mm regions.    As a result, the SED of around a few
hundred AGN are relatively well known but only a handful of the
brightest and low-redshift sources have been observed
simultaneously from $\gamma$-ray to radio (see Wilkes et al. 1999
and reference therein).

The standard SED of AGN, presented in Figure\ref{fig7b}, typically
shows bumps in the optical-UV and IR regions with an inflection
point at $\sim1\mu$m. The near-IR bump generally peaks at
$25-60~\mu$m decreasing rapidly towards the radio region. The
second bump is the Big Blue Bump (BBB) that dominates the
optical-UV emission (1000 - 3000 \AA). The third feature is the
soft X-ray excess (Boller et al. 1996), produced by the steepening
of the X-ray spectra below 1 keV.

A well-established correlation between X-ray and optical emission
in AGN shows  that higher luminosity sources have relatively weak
X-ray emission. A strong correlation was found between X-ray slope
and the radio emission in core dominated sources. The
radio-infrared emission correlation was found only for core
dominated radio-loud AGN. A well established, strong linear
relation was found between radio and IR emission in normal and
starburst galaxies (Mass-Hesse 1992), which is generally
interpreted as a combination of thermal emission from the dust
heated by star formation and non-thermal emission from supernova
remnants.

Typically, the optical and 1 keV region in AGN yields
$a_{ox}\sim-1.3$  . In some AGN the optical spectrum is steeper
than the optical-to- X-ray spectrum which is inconsistent with a
single non-thermal continuum. Since the optical slopes are often
much steeper than -1.3, the optical and X-ray continua are not
necessary part of the same spectral component. In the 2-20 keV
energy range the X-ray spectra of AGN are characterized by a hard
power-law with an energy spectral slope $\alpha_{x}\sim$-1.0.
Below 1 keV $\alpha_{x}$ increases indicating a soft X-ray excess.

\begin{figure}

\figurenum{6a} \plotone{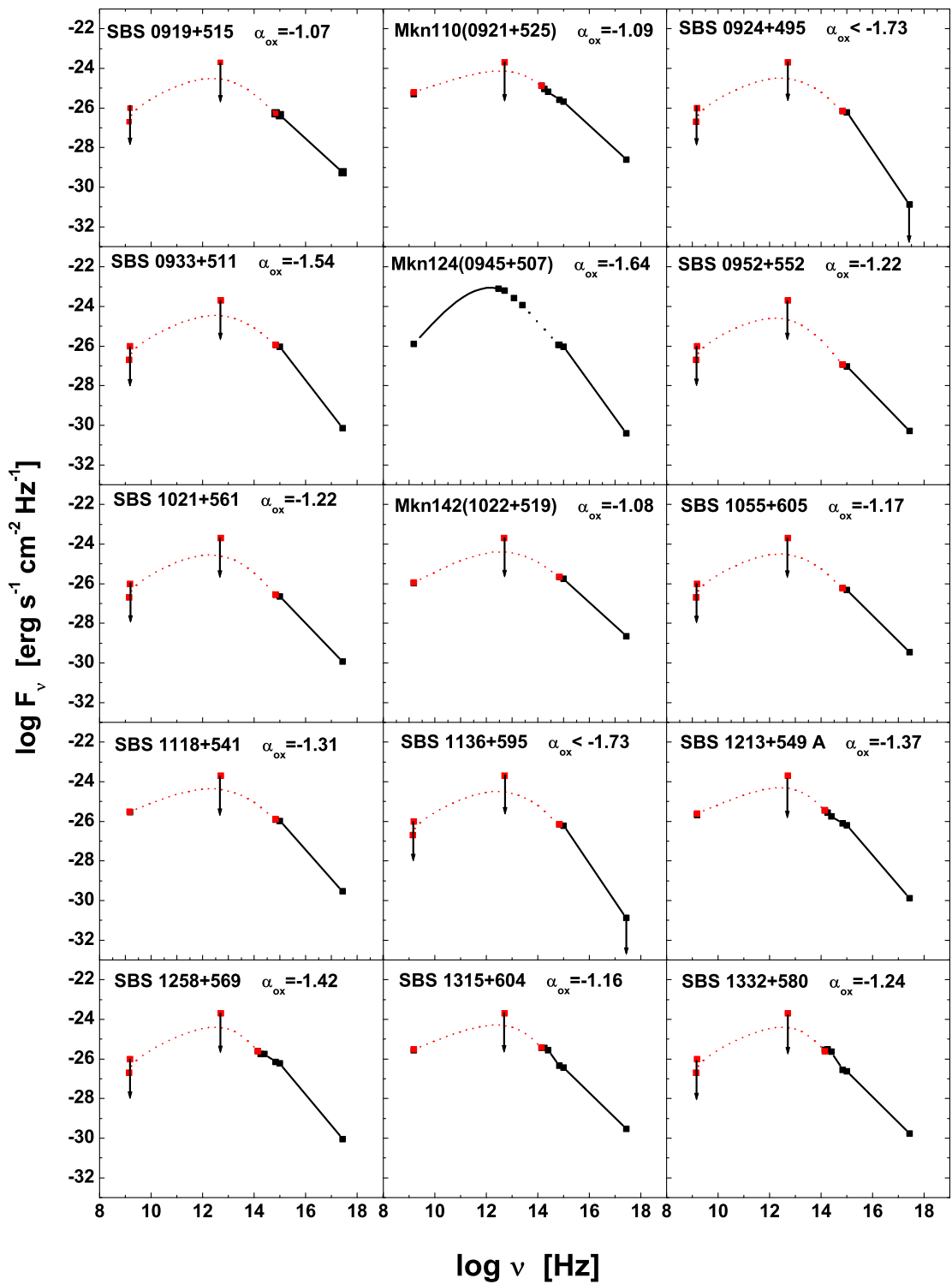} \figcaption[f6a.eps]{Spectral
energy distribution of SBS NLS1s\label{fig6a}. The vertical arrows
show the flux limit detection value for ROSAT, IRAS and FIRST
surveys. The dotted lines show the SED for each individual object
with an estimate of the upper flux limit. The thick solid lines
represent true detections.}

\end{figure}

\begin{figure}

\figurenum{6b} \plotone{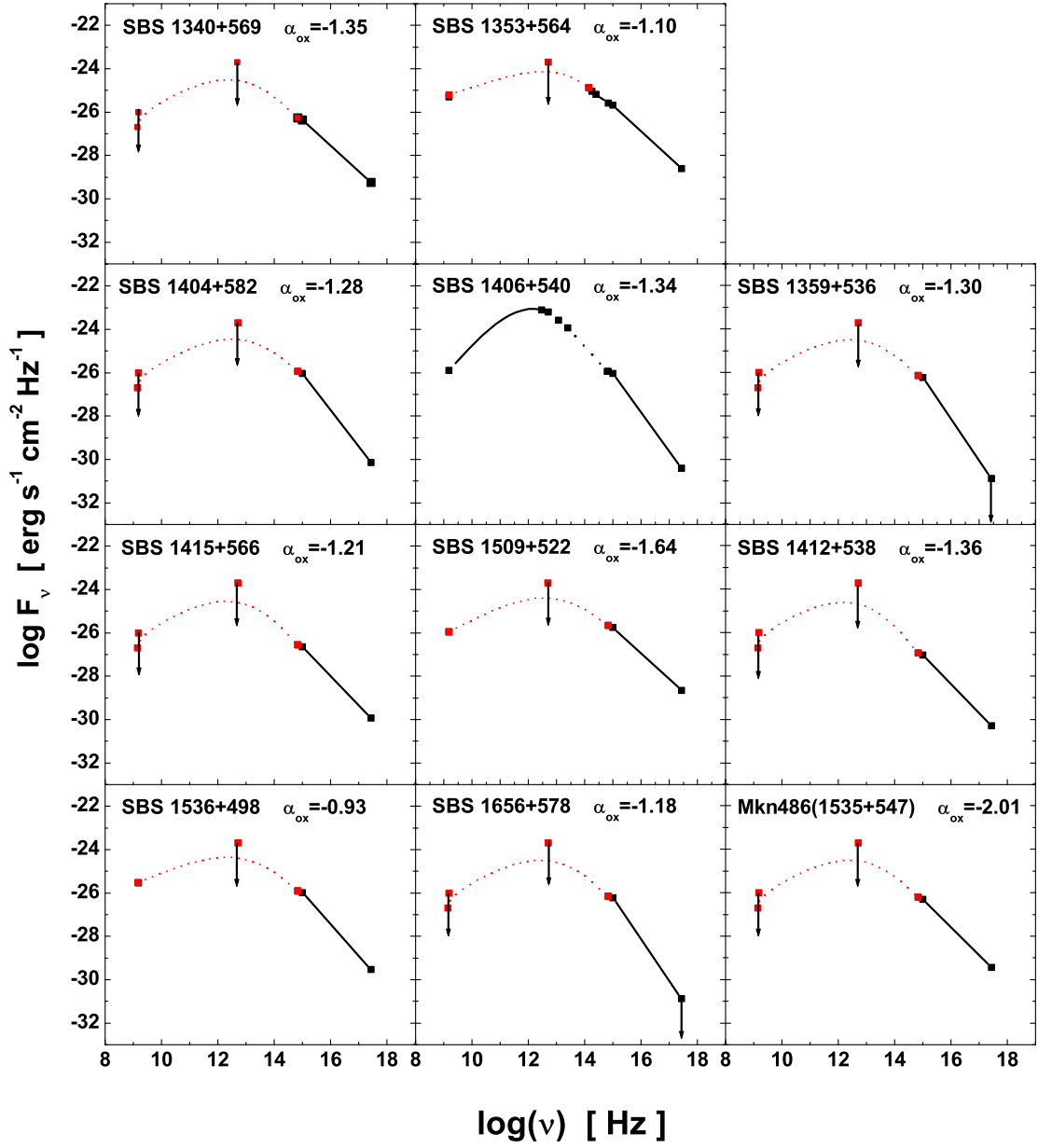} \figcaption[f6b.eps]{Spectral
energy distribution of SBS NLS1s. Continued.\label{fig6b}}

\end{figure}

\begin{figure}

\figurenum{7a} \plotone{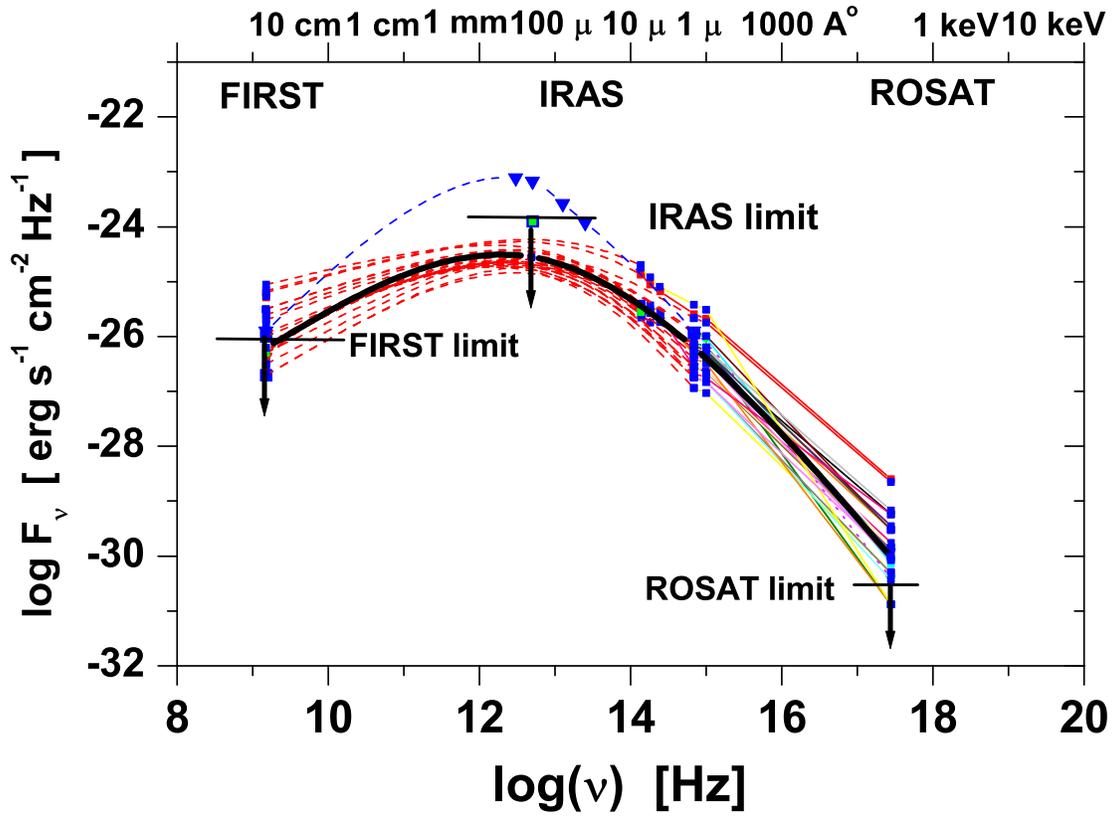} \figcaption[f7a.eps]{Spectral
energy distribution of SBS NLS1s. A wide spread of indices
($-0.93<\alpha_{ox}<2.01$) is observed. The dotted lines show the
SED for each individual object. The thick solid line represents
the mean overall SED. Assuming a power-law fit for the soft X-ray
-to-optical region we get a mean value of $\alpha_{ox}$ =-1.33.
\label{fig7a}}

\end{figure}

\begin{figure}

\figurenum{7b} \plotone{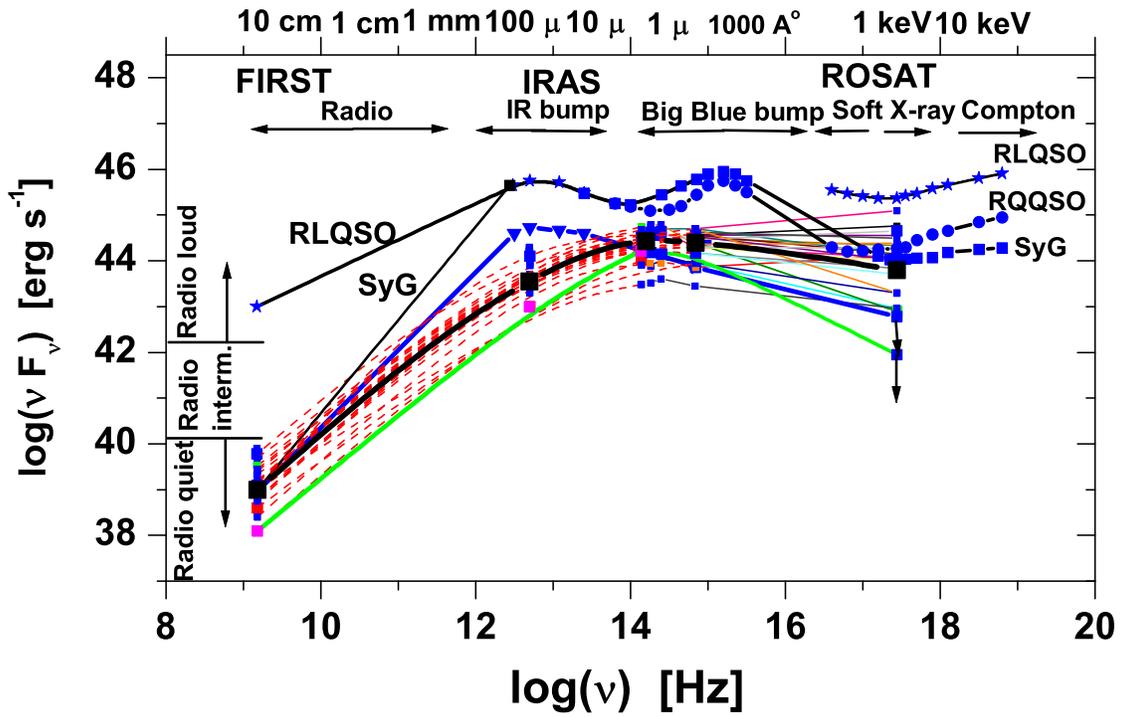} \figcaption[f7b.eps]{Spectral
energy distribution of SBS NLS1s. The dotted lines show the SED
for each individual object. In this case, the SEDs of typical
radio-loud QSOs, radio-quiet QSOs and Seyfert galaxies are shown
for comparison.\label{fig7b}}

\end{figure}

We have derived the spectral energy distributions for SBS NLS1s
from the soft X-ray to radio, using the monochromatic luminosities
$\nu \, L_{\nu}$, as well as the upper detection limit estimates
to study their multiwavelength properties and to investigate
whether the continuum emission is originated by the same
mechanisms. In Fig \ref{fig6a} and \ref{fig6b} the SED for each
object of our SBS NLS1s sample is shown. The composite SED is
shown in Figure\ref{fig7a} and \ref{fig7b}. In the infrared and
radio regions, our data are insufficient to make a reliable
conclusion about their SED behavior. The use of upper detection
limits in IR or radio regions allow us to estimate a probable SED.
The SED of typical radio-loud QSOs (RL QSO), radio-quiet QSOs (RQ
QSO) and Seyfert galaxies (SyG) are shown in Figure\ref{fig7b} for
comparison.

We calculated $\alpha_{ox}$, the slope of the power law defined by
the rest-frame flux densities at 3000 \AA~ and 2 keV for all SBS
NLS1s. Results are presented in the last column of Table \ref{t5}.
A wide spread of indices ($-0.93<\alpha_{ox}<2.01$) are observed
for SBS NLS1s (see Figure\ref{fig7a} and \ref{fig7b}). Sources not
detected in soft X-ray show a very steep slope
$-1.6<\alpha_{ox}<-2.1$. This feature, along with the large
luminosities involved ($\, \log \,L(B)$), partially agree with the
characteristics of weak soft X-ray sources defined by Brand et al.
(2000).

Three objects in our sample, Mkn 110, Mkn 124 and Mkn 486, were
also studied by Brandt et al. (2000), who found $\alpha_{ox}$
equal to -1.41, -1.34 and -2.45, respectively. A large difference
is found as compared to our data, which could be due to
variability or a calculation difference. A possible explanation is
that it may arise from flux calculations in the optical or X-rays
(or both). A difference of 0.5 -1 in $M_{B}$, results in a
difference in $\alpha_{ox}$ of 0.2-0.4. For example, for Mkn 486
we have only an upper detection limit estimate $\alpha_{ox}<
-2.01$, which is in agreement with Gallagher et al. (2001) result
of $\alpha_{ox}< -2.03$. We think that this discrepancy is not a
result of $\alpha_{ox}$ intrinsic variability, but it is more
likely a result of differences in optical and 2 keV monochromatic
flux density calculations.

We have used our $\alpha_{ox}$ value to examine correlations with
other parameters tabulated in Tables 3 to 5. In these correlation
analysis, we have excluded sources which were not detected in soft
X-ray.

\begin{figure}

\figurenum{8} \plotone{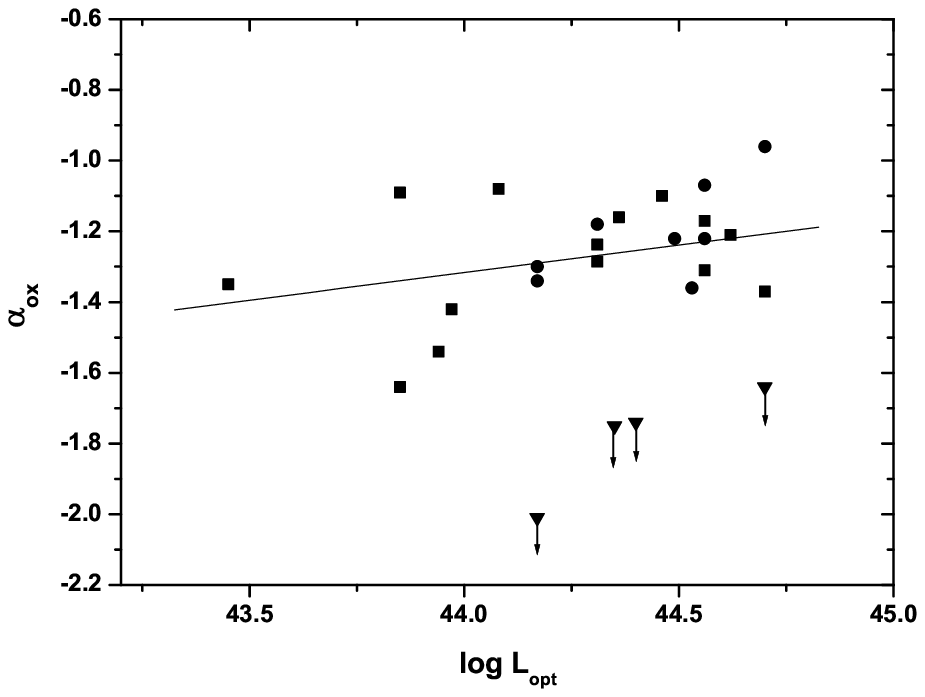} \figcaption[f8.eps]{$\alpha_{ox}$
versus $L_{opt}$ for the SBS NLS1s. A weak correlation
($\alpha_{ox}=0.16 \log L_{opt}-8.2$) can be observed for bright
($B\le17.5$, $z\le0.16$, $\blacksquare$) sources. The solid line
represents the best fit. Weak sources ($B>17.5$,
{\LARGE$\bullet$}) are also shown, as well as upper limits
($\blacktriangledown$).\label{fig8}}

\end{figure}

We find a weak correlation between $\alpha_{ox}$ and $\, \log
\,L_{opt}$ (see Figure\ref{fig8})for bright ($B\le17.5$,
$z\le0.16$) SBS NLS1s, in the form :

\begin{equation}
\alpha_{ox}\,=\,0.16(\pm0.13)\, \log \, L_{opt}\,-\, 8.5(\pm3.7),
\end{equation}

with r=0.35 and  N=14. A similar result was obtained by Wilkes et
al. (1994) and Wilkes (1999): $\alpha_{ox}$\,=\,0.11($\pm$0.02)\,
$\log \,L_{opt}$ - constant.

The existence of a relation between $\alpha_{ox}$ and $\, \log \,
L_{opt}$ has been discussed by a number of authors (Wilkes 1999
and references therein). Wilkes (1999) accepts a well-established
correlation between X-ray and optical emission in AGN, such that
higher luminosity sources have relatively weak X-ray emission,
while other authors suggest that the relation is an artifact
caused by an intrinsic dispersion in $L_{opt}$ which is much
larger than that in $L_{x}$. La Franca et al. (1995) suggest that
the relation is linear and that the discrepancy arise from
photometric errors dominating the dispersion relation. However,
Wilkes (1999) points out that in reality errors in $L_{opt}$ are
significantly smaller than the dispersion. For SBS NLS1s this
relation in general is weak, but it seems even stronger than
Wilkes' result.

We have derived the mean overall SED for the SBS NLS1s (thick
solid line in Figure\ref{fig7a} and \ref{fig7b}). Assuming a
power-law fit for the soft X-ray -to-optical region we get a mean
value of $\alpha_{ox}$ =-1.33.

\section{Discussion}

A complete sample of  QSOs and Sy galaxies of the SBS
survey,($B\le17.5$) of $\sim250$ was isolated and detailed
investigated both spectroscopically and photometrically. This
sample has been used in this work to investigate the
multiwavelength properties of 26 SBS NLS1s.

An independent and direct evidence that the ROSAT sample of AGN
contains only the brightest soft X-ray sources, comes from
comparing SBS and ROSAT samples of AGN (Boller et al. 1992, 1996,
Shwope et al. 2000). The compilation of the AGN ROSAT sample was
done in the same way as the SBS optical survey: preliminary
identification of the optical counterparts of X-ray sources on the
objective prism spectra. A huge amount of optically bright, but
soft X-ray relatively weak AGN, including NLS1s were not included
in ROSAT sample of AGN. Obviously, objects which are below the
detection limit of the ROSAT survey (four of which are studied in
this paper) were not investigated either.

{\bf Soft X-ray (ROSAT 0.05-2.4 keV) region}

SBS NLS1s detected in soft X-rays (22/26) are strong or relatively
strong ($42.8<\, \log \,L_{x}\,<\,45.4$) sources. Most of them,
18/26 objects (69\%), are ROSAT bright sources with count rates
$>$0.05 cts $s^{-1}$, while four objects (15\%) are identified as
RASS faint sources with count rates $<$0.05 cts\,s$^{-1}$. Twenty
NLS1s have soft X-ray fluxes below 6 $\times
10^{-12}$\,erg\,s$^{-1}\,$cm$^{-2}$, and only Mkn 110 and Mkn 142
exceed this value, 23 and 20
$\times10^{-12}$\,erg\,s$^{-1}$\,cm$^{-2}$, respectively.

As might be found from Table \ref{t5}, the mean energy conversion
factor for objects in the sky area covered by SBS is
$\overline{ECF}\sim0.8$. From this value and following Bade et al.
(1995), the count rate limit in ROSAT band (0.05-2.4 keV) of 0.01
cts $s^{-1}$ corresponds to the ROSAT detection limit of $1.25
\times 10^{-13}$\,erg\,$cm^{-2}$\, $s^{-1}$. The corresponding
luminosity upper limits for non detected sources in soft X-ray
band are $\, \log \,L_{x}<42.9$ for SBS 0924+495 and SBS 1136+595,
$\, \log \,L_{x}<43.4$ for SBS 1509+522 and $\, \log \,L_{x}<41.9$
for Mkn 486 (SBS 1535+547).

In spite of the absence of strong X-ray emission, SBS NLS1s not
detected by ROSAT share all the typical properties of these kind
of objects. This is puzzling since they could be intrinsically
weak X-ray sources or highly absorbed.

Practically all known X-ray properties of NLS1s are based on
samples drawn from the RASS Bright Source Catalogue. In those
works no evidence was found for significant internal absorption on
the majority of AGN (Boller et al. 1996). Similar results were
obtained by Grupe et al. (1998), who studied a sample of 76 ROSAT
Bright AGN (partially crossed with the Boller et al. sample).
These results suggest that an explanation for the lack of soft
X-ray detection in several SBS NLS1s could be that they are
intrinsically weak X-ray sources below the detection limit of
ROSAT survey.

However, recent studies by Brandt et al. (2000) and Gallagher et
al. (2001) contradict this explanation. Brandt et al. (2000) have
identified and studied the soft X-ray weak AGN population in the
BQX optically selected sample of Boroson \& Green (1992) of all PG
QSO with $z<0.5$, including a few NLS1s. They found that 11\% of
BQX QSOs are weak soft X-ray sources, which is similar to our
results. ASCA observation of soft X-ray weak NLS1s, show that they
are heavily X-ray absorbed objects. In general ASCA observations
support the intrinsic absorption scenario for explaining soft
X-ray weakness in AGN, particulary in the NLS1 Mkn 486. Weak soft
X-ray SBS NLS1s are good candidates for ASCA observations.

One object in our sample, Mkn 486 has clear observational evidence
of the presence of high intrinsic absorption. As was shown by
Gallagher et al. (2001), Mkn 486 was not detected in the soft
X-ray 0.05-2.0 keV band, only in the 0.1- 5 keV band. The source
shows clear signs of high intrinsic absorption by $N_H=1.2 \times
10^{23} cm^{-2}$, with only partial covering of the power-law
continuum. The hard X-ray source has an ASCA photon index
$\Gamma$=$2.02\pm0.93$. The central emission source is totally
covered by neutral gas, therefore the low energy photons are
completely absorbed there. The object has high optical continuum
polarization (P=2.5\%, Berman et al. 1990).

The SBS NLS1s sample is based on the identification of both RASS
Bright and Faint sources. If we suppose that non detected SBS
NLS1s in X-ray are objects like Mkn 486, it may mean that at least
15\% of SBS NLS1s may have hidden AGN in their central zone, and
would be detected in the optical (UVX, Markarian) and perhaps in
other wavebands, but not in the soft X-ray region. The existence
of a hidden AGN in the central zone of these galaxies may explain
the weakness or absence of soft X-ray emission.

There is a number of arguments which has been used against a
hidden AGN in NLS1s. One of the strongest arguments used is the
steepness of the soft X-ray spectrum, which is interpreted as
evidence that the central source is seen directly. The SED of
NLS1s does not support the picture of hidden AGN (the SED of NLS1s
is presume to be closer to that of Sy1 than to Sy2, the later been
more likely a hidden AGN). We recall that the arguments against a
hidden AGN are based on studies of the brightest X-ray sources.

In the SBS sample of NLS1s, non detected sources have a
$f_{x}/f_{opt} <-4.7$, which means that they are 10 to 50 times
less luminous in soft X-rays relative to the bulk of NLS1s or
ROSAT AGN. SBS NLS1s which were not detected in X-ray were also
not detected neither as IRAS sources, nor as radio sources, except
Mkn 486 which is the only radio-quiet AGN observed by Neugebauer
\& Mattews (1999).

{\bf Far-Infrared (IRAS $12-100 \mu$) region}

All SBS NLS1s except Mkn 124 (SBS 0945+507) were not detected as
IRAS sources. One of the natural explanations is that it is due to
the flux limitation of IRAS survey (see the redshift distribution
of IRAS sources, Saunders et al. 2000), which is perhaps true for
high and intermediate redshift objects. Among the nearby objects
with $z<$0.055, only Mkn 124, was detected by IRAS.  This object
is a Luminous Infrared Galaxy
(LIG,{$\log\,[L_{FIR}/L_\odot]=11.07$}) and also the weakest soft
X-ray source in our sample. Mkn 124 is a rare case, which combines
high far-infrared luminosity ($\log \, L_{FIR}$=44.65) with low
X-ray luminosity ($\, \log \,L_{x}$=42.8). In Boller et al.
(1992), none of the objects combines a high far-infrared with a
low X-ray luminosities.

The limiting flux in the FIR is 0.5 Jy at $60 \mu$m and 0.8-1.0 Jy
at 100 $\mu$m (IRAS PSC, Boller et al. 1996), and 2.5 times less
for IRAS faint sources, 0.2 Jy at 60 $\mu$m and 0.35-0.4 Jy at 100
$\mu$m for (IRAS FSC), which corresponds to an IRAS flux limit
detection $F_{FIR}=2.6 \times 10^{-11}~erg~ s^{-1}~ cm^{-2}$ and
$1.1 \times 10^{-11}~ erg~ s^{-1} ~cm^{-2}$, respectively.

For nearby objects $z<0.05$, IRAS flux limit detection corresponds
to {$\, \log \,[L_{FIR}/L_\odot]<10.5$}, which means that non
detected nearby SBS NLS1s are not LIGs.

A similar estimate might be obtained for the redshift range
$0.05<z<0.21$ (only two objects have $z>0.21$). This redshift
range corresponds to the luminosity range {$10.5>\, \log
\,[L_{FIR}/L_\odot]<11.7$}. If the objects are Ultraluminous
Infrared galaxies (ULIGs) ($\, \log \,[L_{FIR}/L_\odot]>12.0$), we
expect that all of them should be detected. If they are LIG, most
of them should also be detected. The absence of detection means
that they are not ULIG  and that a significant part of them may
not be LIG either.

According to Boller et al. (1992), IRAS galaxies are in general
faint X-ray emitters. They show that X-ray and IR luminosities are
correlated. LIGs tend to be also X-ray luminous (see their fig.
6). Soft X-ray luminous sources in our sample of NLS1s do not tend
to be luminous in the infrared. The X-ray luminosity range of SBS
NLS1s is $42.8<\, \log \,L_x<45.5$. Following Boller et al. (1992)
we would expect that they may have IR luminosities in the same
range, i.e. $43.5<\, \log \,L(IR)<45.5$. As was shown above, if
the SBS NLS1s have such luminosities, they should have been
detected by IRAS. We conclude, that most of the SBS NLS1s are
luminous X-ray sources and weak FIR sources. These objects would
be located in the upper left part of fig. 6 of Boller et al.
(1992) and may drastically change their correlation.    NLS1s were
found as luminous infrared emitters also by Halpern \& Oke (1987)
and Lipari (1994) who have investigated exclusively the
ultraluminous and luminous infrared sources with extreme FeII
emission.

Keel et al. (1994) investigated the correlations between
emission-line properties and IR parameters for 221 warm IRAS AGN
from the Point Source catalogue. They found that optical emission
line properties are correlated with the IR emission which is
consistent with the unified model. Steeper IR spectra is
statistically associated with the strongest FeII emission. FeII
emission is believed to be produced in very dense zones,
$n_e\sim10^{10}~ cm^{-3}$, within the broad line region.
Therefore, the FeII -- infrared connection poses an important
challenge to the unified model.

In this scenario, a screen of surrounding material absorbs optical
light which is reradiated in the IR. The torus blocks escaping
radiation except in the direction along its symmetry axis which is
believed to coincide with the axis of the central machine (the
radio axis). The emission intercepted by an obscuring thick torus
should be reradiated in the infrared. The warmest IR spectra would
be associated with torii viewed face-on, so that their inner
region are fully exposed along the line of sight, while an edge-on
torus would be cooler because we preferentially see the outer
regions of the torus.

If strongest FeII emission is statistically associated with
steeper IR spectra, we would expect, that SBS NLS1s, would have
strong IR emission, since more than half are either strong or very
strong FeII emitters. This is not the case. Therefore, the
proposed relation between optical-to-IR spectral slope and FeII
emission is more likely to be an artifact of sample selection.

Moran et al. (1996) obtained accurate spectroscopic classification
of the Boller et al. (1992) sample of IRAS sources detected by
ROSAT. They find a correlation between $L_{x}$ and $L_{FIR}$ for
luminous IRAS/ROSAT objects. However, this correlation is still
under debate. It remains unclear whether the IR emission in AGN is
primarily thermal and whether both X-ray and IR emission arise
from a common non-thermal source. The inspection of 20 NLS1s
objects in table \ref{t4} of Moran et al. (1996), shows that all
of them without exception are LIGs or ULIGs. The cross correlation
of two bright source catalogues (IRAS and ROSAT BSC), resulted in
the selection of the brightest objects both in X-ray and in
infrared. Moran et al. (1996) conclude that NLS1s are very
luminous soft X-ray and FIR emitters. Nevertheless, this is
another example of sample dependent results.

>From the data presented by Boller et al. (1992) and Moran et al.
(1996), the proportion of NLS1s yields $\sim 9.5\%$ of the
luminous IRAS/ROSAT sources and $\sim17\%$ among the luminous
IRAS/ROSAT Sy1s.

In the SBS sample of NLS1s, there is only one object which is LIG.
The others are not ULIG and most of them may not be LIG (see
Figure\ref{fig6a}, \ref{fig6b}, \ref{fig7a} and \ref{fig7b}). The
SBS NLS1s not detected by IRAS, may have SEDs without a FIR peak.
This is important for NLS1s physics. Therefore, it is not clear
how representative previous results are on the X-ray and/or IR
properties of NLS1s parent population. The IR bump is in close
connection with the BBB bump, because the IR part of the BBB
emission seems to be reprocessed by a circum-nuclear dust torus
(Krolik 1996). Rodrigues- Pascual et al. (1997) discussed the
hypothesis of the absence of the BLR in some NLS1s, as well as the
possibility that the line emitting material is optically thin. If
these are correct, then the absence of an IR bump in the majority
of SBS NLS1s and the weakness of X-ray radiation of $\sim15$\% of
them might be used as evidence of the absence of a BLR in these
NLS1s (if they are intrinsic properties). Furthermore, no evidence
of BBB was found in the radio-loud NLS1 PKS 2004-447 by Oshlack et
al. (2001). We suggest that almost all SBS NLSy1s may not have the
FIR bump, nor a BLR, but may have a BBB bump based on their SEDs.

{\bf Radio (1.4 GHz) region}

Nine SBS objects (33\%) are detected  at 1.4 GHz, with radio
fluxes between 1 and 6 mJy. The corresponding radio luminosities
are $38.0 < \, \log \, L_{R} < 40.0$, similar to Mkn galaxies and
characteristic of ``normal galaxies'' (Bicay et al. 1995).

The radio region is still unknown for radio-quiet AGN, especially
for NLS1s (Wilkes 1999). A sample of seven NLS1s was studied in
the radio by Ulvestad et al. (1995). They found that they are
faint sources with radio-power from $10^{20}$ to $10^{23}$ W
Hz$^{-1}$, similar to that found for Mkn galaxies by Bicay et al.
(1995). Moran (2000) investigated  24 NLS1s at 20 and 3.6 cm using
the VLA. Most were found to be unresolved sources at $\sim 1
\arcsec$, with generally  steep ($\alpha\sim1.1-1.2$) radio
spectra; they are generally radio-quiet objects, with only four
being radio-loud: PKS 0558-504 (Remillard et al. 1986, Siebert et
al. 1999), RGB J0044+193 (Siebert et al. 1999), J0134.2-4258
(Grupe et al. 2000) and PKS2004-447 (Oshlack et al. 2001). Siebert
et al. (1999) found no clear differences between radio-loud and
radio-quiet NLS1s.

The importance of radio loud NLS1s lies in the fact that the
presence of radio emission, and an associated relativistic jet,
might give us an independent light on the orientation issue. The
discovery of radio-loud NLS1s indicate that the observational
definition of NLS1s requires refinement.

Only 10 SBS NLS1s have radio data and are radio-quiet sources.

{\bf FeII emission}

One of the general properties of SBS NLS1s is the presence of
strong FeII emission, usually quantified by $R(4570)$. In our
sample it shows a wide range of values, between 0.1 and 1.6. More
than half of them (52\%) are strong FeII emitters, $R(4570) \,>\,
1$. The mean $\overline{FeII\lambda4570}$ flux value obtained is
38$\pm$4, and the mean $\overline{R(4570)}$ value is
0.90$\pm$0.44.

Two basic correlations for NLS1s are known, both related to the
strength of FeII: FWHM of H$\beta$ and soft X-ray photon index
strongly correlated with $FeII/H\beta$ strength. These
correlations suggest, that the strength of FeII emission is
closely related to the soft X-ray emission. We obtain a similar
result for SBS NLS1s (see Figure\ref{fig3}). Weak FeII emitters
have always large FWHM but strong FeII emitters have either large
or small FWHM. Marziani et al. (2001) show that the R(4570) ratio
versus $FWHM(H\beta)$ may define an AGN ``main sequence'', with
NLS1s located at the strong end of R(4570).

We were not able to investigate the correlation between FeII
emission and radio properties of SBS NLS1s because of the scarce
number of sources detected at 1.4 GHz. The correlation of the
radio spectrum with the strength of FeII emission was discussed by
Boroson \& Green (1992). They found that radio-quiet QSOs tend to
have large value of EW FeII and small value of [OIII] (fig. 2 of
Boroson \& Green). Strong FeII and weak [OIII] tend to have narrow
lines (NLS1s/NLQSOs) and objects with weak FeII and strong [OIII]
predominantly are steep-spectrum radio-loud QSOs.

{\bf Surface density}

The data presented in this paper allows us to make the first rough
estimate of the surface density of NLS1s. The objects with
$B\le17.5$ might be considered as members of a complete sample of
SBS NLS1s. The lower limit value of the surface density of SBS
NLS1s is about 0.015 per $deg.^2$ ($B<17.5, z<0.16$). The
proportion of NLS1s among the complete sample of SBS Sy1s in the
sky area of about $1000~ deg^2$ consists of about $\sim40\%$. More
precise data, together with the results of high-resolution
spectroscopy and photometry for the SBS NLS1s complete sample will
be presented in a forthcoming paper.

The absolute magnitude of SBS NLS1s are ranged from $-19.9
>M_{B}>-23.0$, optical luminosities between 43.5 $ < \, \log \,
L(B)\, <$ 44.7, soft X-ray (0.05-2.4 keV) luminosities between
42.8 $< \, \log \,L_{x} \,<$ 45.4 and radio luminosities at 1.4
GHz between $38.0< \, \log \,L_{R} <40.0$.

\section{Conclusions}

A multiwavelength investigation of the homogeneously selected
sample of 26 SBS NLSy1s is presented. Basic data (accurate
coordinate, magnitudes, redshifts, identification of optical
objects with X-ray, IR and radio sources), together with data
derived from photometric magnitudes, fluxes and spectroscopy (EW
and FWHM of H$\beta$, [OIII]$\lambda$5007, FeII$\lambda$4570,
[OIII]$\lambda$5007/H$\beta$, FeII$\lambda$4570/H$\beta$, etc.)
are presented. This allowed us to construct the SED for a complete
sample of NSL1s.

Our main conclusions are:

\begin{enumerate}

\item They are mainly  strong or moderately strong ($\, \log
\,L_{x}$= 42.8-45.4) soft X-ray sources. Non soft x-ray detected
sources have a $f_{x}/f_{opt} <-4.7$, which means that they are 10
to 50 times less luminous in soft X-rays relative to the bulk of
NLS1s or ROSAT AGN. Four objects not detected in soft X-ray band
are optically luminous $\, \log \,L(B)>44.2$.

\item SBS NLS1s do not tend to be luminous in the infrared. Most
of them (92\%) were not detected by IRAS. They are not
ultraluminous infrared galaxies, and most of them are not luminous
infrared galaxies. They are predominantly weak infrared sources.
Among them the only LIG galaxy is Mkn 124, which is a rare NLS1s
which combines a high FIR luminosity with a low X-ray luminosity.

\item All SBS NLS1s which have detections in the radio are
radio-quiet objects, $\, \log \,L_{R}<40.0$.

\item All SBS NLS1s show FeII emission, with more than half of
them (52\%) being moderately strong FeII emitters,
FeII$\lambda$4570/H$\beta >$1.

\item  Like in most AGN, a linear correlation between the X-ray
luminosity and optical luminosity is found for the SBS NLS1s, $\,
\log \,L_{x}=1.34~ \, \log \,L_{B}$ - 15.3. Also, an
anticorrelation between FWHM of ${H\beta}$ and the ratio of
FeII$\lambda$4570/H$\beta$ is obtained.

\item A weak correlation is found between $\alpha_{ox}$ index and
optical luminosity, $\alpha_{ox}=0.16~\, \log \,L_{opt}$ - 8.5.
The mean value of the power-law slope $\alpha_{ox}$ for the SBS
NLS1s -1.33.

\item The first estimate of the surface density of NLS1s is
obtained: 0.015 per $deg^2$ ($B\le17.5, z\le0.16$). The proportion
of NLS1s among the Sy1s in the complete sample of SBS AGN
comprises 40\%.

\item A comparison of the luminosities in the FIR, optical and
soft X-ray continuum band leads to the conclusion, that the SED of
SBS NLS1s may differ from those studied before. The traditional
peak in FIR which is seen in most AGN is not seen in many of these
objects. Most of the SBS NLS1s which were not detected by IRAS and
may not have the FIR bump. Therefore, the SBS NLS1s are excellent
candidates for SIRTF observations, since they were not detected by
ISO.

\item The absence of the FIR bump in the majority of the SBS NLS1s
coupled with the suggestion that they could possess a BBB and the
weakness of X-ray radiation in some of them (if they are intrinsic
properties) may argue against the presence of a BLR in these
objects.

\item The SBS sample of NLS1s, in spite of their small number,
consist of a physically heterogenous group of objects. They do not
have physical properties identical as the NLS1s studied so far.
The observational properties of NLS1s may not depend predominantly
on their orientation but on individual evolutionary history. Among
SBS NLS1s, we found  controversial objects which do not
necessarily fulfill the  ``main'' properties of NLS1s: At least
15\% of SBS NLS1s may have a hidden AGN in their central zone, and
therefore may be objects highly absorbed, like Mkn 486. They are
weak soft X-ray sources, but optically luminous objects. Objects
which are strong soft X-ray sources, but weak FIR sources and vice
versa. Objects with moderately strong FeII emission, but weak soft
X-ray emission and vice versa.

\end{enumerate}

In summary, accordingly with the data presented in the literature
and the observational properties of SBS NLS1s presented in this
paper, we can say that only the tip of the iceberg was studied,
i.e. the most luminous sources. To obtain a more realistic
picture, one needs to investigate the X-ray and IR weak NLS1s,
which outnumbers (200-300) the brightest ones. The detection of
weak soft X-ray NLS1s, radio loud NLS1s, weak FeII emission NLS1s,
as well as the presence of objects with the FWHM of H$\beta$ less
than 2000 km $s^{-1}$, but [OIII]/H$\beta>3$ make the NLS1s
definition very smooth and artificial. Therefore, this definition
needs to be refined. NLS1s may not be formally a distinct class of
objects. The velocity cut FWHM of H$\beta<2000$~ km~ $s^{-1}$, and
[OIII]/H$\beta<3$, automatically and artificially isolated this
``narrow'' group of AGN, that are called NLS1s. We support that in
general, there is a continuation of all properties of NLS1s to the
class of more luminous objects with the luminosities
$M_{B}>-23.0$, NLQSOs, which are the bright cousins of NLS1s; as
well as, a continuous transition of all properties between NLS1s
and classical BLS1s.

\section{Acknowledgements}

J. A. Stepanian acknowledges financial support from CONACYT
(C\'atedra Patrimonial II  EX-000287).  E. Ben{\'\i}tez and I.
Cruz-Gonz\'alez acknowledge support from grant ES118601 from
DGAPA-UNAM, which includes also T. Verdugo student scholarship.
This research has made use of NASA/IPAC Extragalactic Database
(NED), which is operated by the Jet Propulsion Laboratory,
California Institute of Technology, under contract with the
National Aeronautics and Space Administration. Finally, we want to
thank our anonymous referee for very useful comments which helped
to improved this work.

\clearpage

\begin{deluxetable}{llllllllcccll}

\tablecolumns{13} \rotate \tabletypesize{\tiny} \tablewidth{0pc}
\tablecaption{Identification and the basic data \label{t1}}
\tablehead{ \colhead{SBS} & \colhead{R.A.} & \colhead{Dec.} &
\multicolumn{5}{c}{Identification} & \colhead{d*d} &
\multicolumn{2}{c}{FeII} & \colhead{Other} & \colhead{Ref.}
\\ \cline{4-11} \colhead{Design.} & \colhead{2000} & \colhead{2000} & \colhead{
1RXSJ} & \colhead{o/x} & \colhead{radio} & \colhead{o/r} &
\colhead{NIR} & \colhead{\arcsec} & \colhead{$\lambda4570$} &
\colhead{$\lambda5100$} & \colhead{name}\\
\colhead{(1)} & \colhead{(2)} & \colhead{(3)} & \colhead{ (4)} &
\colhead{(5)} & \colhead{(6)} & \colhead{(7)} & \colhead{(8)} &
\colhead{(9)} & \colhead{(10)} & \colhead{(11)} & \colhead{(12)}
& \colhead{(13)} }

\startdata

0919+515                  &09 22 47.17& +51 20 38.6& 092246.4+512046& 10& \nodata &\nodata & \nodata&   8   & vs & vs &     & [1,2]  \\
0921+525\tablenotemark{a} &09 25 12.89& +52 17 10.9& 092512.3+521716&  7\tablenotemark{b}& FIRST & $<1$ &\nodata &25*12 & w & w  & Mkn 110 & [3,4,17,19] \\
0924+495                  &09 28 01.22& +49 18 16.9& Non ROSAT      & \nodata & \nodata& \nodata &\nodata   & 10 & vs& vs &         & [6,7]            \\
0933+511                  &09 36 43.12& +50 52 49.6& 093642.6+505249&  5& \nodata &\nodata &\nodata & 24*12 & s & s &         & [7,17]                 \\
0945+507                  &09 48 42.66& +50 29 32.2& 094841.6+502926& 12& FIRST   &$<1$    &\nodata & 12    & s & s & Mkn 124 & [4,5,8,9,17,19]        \\
0952+552                  &09 56 13.38& +54 59 05.5& 095613.3+545904&  2& \nodata &\nodata &\nodata & 5     & w & w &         & [18]                   \\
1021+561                  &10 24 34.70& +55 56 25.7& 102435.5+555644& 19& \nodata &\nodata &\nodata & 8     & vs& vs&         & [6,17]                      \\
1022+519                  &10 25 31.21& +51 40 34.8& 102531.2+514039&  4\tablenotemark{b}& FIRST & $<1$ &\nodata & 18*14 & vs & vs & Mkn 142 & [10,17,19]   \\
1055+605                  &10 58 30.13& +60 16 00.5& 105830.1+601602&  2\tablenotemark{b}& \nodata&\nodata&\nodata& 8 & vs & vs & RBS 925 & [10,17,20]      \\
1118+541\tablenotemark{a} &11 21 08.55& +53 51 20.7& 112109.9+535125& 12\tablenotemark{b}& FIRST & $<1$ &\nodata & 12 & vs & vs &RBS 971 & [12,17,19,20]    \\
1136+595                  &11 39 00.48& +59 13 46.7& Non ROSAT      & \nodata & \nodata  & \nodata & \nodata& 14 & vs& s &    &                  \\
1213+549A\tablenotemark{a}&12 15 49.39& +54 42 24.1& 121549.3+544227& 3& FIRST  & $<1$   & 2MASS & 10*8 & s & s &  & [12,13,14,17,19]            \\
1258+569                  &13 00 52.24& +56 41 06.9& 130052.9+564101& 8&\nodata &\nodata & 2MASS & 12   & s & s &  & [17]                        \\
1315+604                  &13 17 50.26& +60 10 40.7& 131750.4+601047& 6& NVSS   &16\tablenotemark{d}&2MASS&15*10& w & w &  & [17]                \\
1332+580                  &13 34 35.25& +57 50 16.3& 133434.3+575019& 8&\nodata &\nodata & 2MASS & 9 & vs& vs&   & [17]                          \\
1340+569\tablenotemark{a} &13 42 10.23& +56 42 11.6& 134210.9+564219& 8&\nodata &\nodata & 2MASS & 14& w & w &   & [17]                          \\
1353+564                  &13 55 16.50& +56 12 44.5& 135515.9+561244& 5\tablenotemark{b}&FIRST  &$<1$  & 2MASS   & 9 & vs& s &   & [15,17,19]    \\
1359+536                  &14 00 50.18& +53 24 24.8& 140048.5+532423&15&\nodata&\nodata &\nodata & 6     & w & w &       & [17]                  \\
1404+582                  &14 06 04.73& +58 00 41.3& 140606.1+580045&12&\nodata&\nodata &\nodata & 7     & vs& s &       & [17]                  \\
1406+540                  &14 07 59.19& +53 47 59.8& 140800.2+534815&18&\nodata&\nodata &\nodata & 6     & s & s &       & [17]                  \\
1412+538                  &14 14 19.78& +53 38 15.4& 141419.2+533803&13&\nodata&\nodata &\nodata & 9     & vs& s &       & [17]                  \\
1415+566                  &14 17 00.11& +56 26 00.9& 141700.1+562601& 2\tablenotemark{c}&\nodata &\nodata &\nodata & 11  & s  & s &RBS 1362 & [15,17,20] \\
1509+522                  &15 11 01.91& +52 03 49.7&    Non ROSAT   &  &       &      &        & 9& vs& w &       &                              \\
1535+547\tablenotemark{a} &15 36 38.43& +54 33 33.0&    Non ROSAT   & 5& FIRST & $<1$ &\nodata &15& vs& s &Mkn 486& [5,16,19]                    \\
1536+498                  &15 37 32.42& +49 42 46.4& 153732.5+494249& 3& FIRST & $<1$ &\nodata & 6& s & vs&       & [15,17,19]                   \\
1656+578                  &16 56 55.93& +57 45 48.4& 165656.5+574541& 8&\nodata&\nodata&\nodata& 8& s & w &       & [17]                         \\

\enddata

\tablenotetext{1}{[a] Pair; [b] According to Schwope et al.
(2000); [c] According to Bade et al. (1995); [d] According to
Bauer et al. (2000); 2MASS - 2 Micron ALL Sky Survey.}

[1] Boller et al. (1996); [2] Stephens (1989); [3] Boroson \&
Green (1992); [4] Osterbrock (1977); [5] Veron et al. (2001); [6]
Stepanian et al. (2001); [7] Stepanian et al. (1993); [8] de Grijp
et al. (1992); [9] Keel et al. (1994); [10] Grupe et al. (1999);
[11] Stepanian et al. (2002); [12] Moran et al. (1996); [13]
Martel \& Osterbrock (1994); [14] Stepanian et al. (1991); [15]
Bade et al. (1995); [16] de Robertis (1985); [17] Voges et al.
(1999); [18] Voges et al. (2001); [19] Becker et al. (1995); [20]
Shwope et al. (2000).

\end{deluxetable}

\clearpage

\begin{deluxetable}{lllllll}

\tablecolumns{4} \tabletypesize{\scriptsize} \tablewidth{0pc}
\tablecaption{Misidentified Galaxies \label{t2}} \tablehead{
\colhead{Name of object} & \multicolumn{2}{c}{J2000} &
\multicolumn{2}{c}{B1950} & \colhead{Ref.} \\
\colhead{(1)} & \colhead{(2)} & \colhead{(3)} & \colhead{ (4)} &
\colhead{(5)} & \colhead{(6)} }
\startdata

1RXSJ121549.3+544227 & 12 15 49.4  & +54 42 28   & 12 13 21.6  & +54 59 07   & [1,4,6]   \\
SBS 1213+549A        & 12 15 49.36 & +54 42 24.3 & 12 13 21.64 & +54 59 04.2 &           \\
FIRSTJ121549.4+544223& 12 15 49.45 & +54 42 23.8 &             &             & [2]       \\
IRAS 12134+5459      &             &             & 12 13 25.3  & +54 59 10   & [3,4,5,6] \\
IRAS 12134+5459      &             &             & 12 13 27.7  & +54 59 12   & [3,5,6]   \\
IRAS 12134+5459      & 12 15 52.1  & +54 42 32.5 &             &             & [3,6]     \\
MCG 09-20-133        & 12 15 55.50 & +54 42 32.9 & 12 13 27.83 & +54 59 12.7 & [4,6]     \\
NVSSJ121553+544229   & 12 15 53.04 & +54 42 29.5 &             &             & [1,7]     \\
\enddata

[1] Voges et al. (1999); [2] Becker et al. (1995); [3] Saunders et
al (2000); [4] Condon et al. (1998a); [5] Moran et al. (1996); [6]
Boller et al. (1992); [7] Condon et al. (1998b).

\end{deluxetable}

\clearpage

\begin{deluxetable}{lllllllllllll}

\tablecolumns{13} \tabletypesize{\scriptsize} \tablewidth{0pc}
\tablecaption{The Equivalent widths and FHWM \label{t3}}
\tablehead{ \colhead{SBS} & \multicolumn{2}{c}{H$\alpha$} &
\multicolumn{2}{c}{H$\beta$} & \multicolumn{2}{c}{$H_\gamma$} &
\multicolumn{2}{c}{[OIII]} & \colhead{FeII$\lambda$4570} &
\colhead{FeII$\lambda$5100} & \colhead{H$\beta$} & \colhead{Ref.}
\\ \cline{2-12}
\colhead{Design.} & \colhead{EW} & \colhead{FWHM} & \colhead{EW} &
\colhead{FWHM} & \colhead{EW} & \colhead{FWHM} & \colhead{EW} &
\colhead{FWHM} & \colhead{EW} & \colhead{EW} & \colhead{FWHM} &
\\

\colhead{(1)} & \colhead{(2)} & \colhead{(3)} & \colhead{ (4)} &
\colhead{(5)} & \colhead{(6)} & \colhead{(7)} & \colhead{(8)} &
\colhead{(9)} & \colhead{(10)} & \colhead{(11)} & \colhead{(12)} &
\colhead{(13)}}

\startdata

0919+515                 & \nodata& \nodata & \nodata &  1390  & \nodata & \nodata & \nodata &   1860  & \nodata & \nodata & \nodata   & [2]       \\
0921+525                 & \nodata& \nodata &  145    &  2120  & \nodata & \nodata &   83    & \nodata &   20    & \nodata & 1670-2500 & [3,5]     \\
0924+495                 & \nodata& \nodata &   90    &  1260  &   22    &  2340   &   36    &   840   & \nodata & \nodata & \nodata   &           \\
0933+511                 &   82   &  1290   &   22    &  1190  & \nodata & \nodata &    4    &   500   & \nodata & \nodata & \nodata   &           \\
0945+507\tablenotemark{a}& \nodata&  1645L  &   43    &  1840L & \nodata & \nodata & \nodata &   540   & \nodata & \nodata & 1050-1400 & [5]       \\
                         & \nodata& \nodata &   53    & \nodata& \nodata & \nodata & \nodata & \nodata &   39    &   42    & \nodata   & [9]       \\
                         & \nodata& \nodata & \nodata &  1400  & \nodata & \nodata & \nodata & \nodata & \nodata & \nodata & \nodata   & [4]       \\
0952+552                 &  390   &  2780   &   98    &  1950  &   32    &  2620   &   50    &   550   &   25    & \nodata & \nodata   &           \\
1021+561                 & \nodata& \nodata &   23    &  1670  & \nodata & \nodata &   20    & \nodata & \nodata & \nodata & \nodata   &           \\
1022+519\tablenotemark{a}& \nodata& \nodata & \nodata &  1710  & \nodata & \nodata & \nodata &   270   & \nodata & \nodata & \nodata   & [9]       \\
                         & \nodata&  1335   &   60    &  1370L & \nodata & \nodata & \nodata &   260   & \nodata & \nodata & 1350-1790 & [5]       \\
                         & \nodata& \nodata &   69    &  1620  & \nodata & \nodata &    7    & \nodata &   74    & \nodata & \nodata   & [3]       \\
1055+605\tablenotemark{a}& \nodata& \nodata & \nodata &  1880  & \nodata & \nodata & \nodata &   470   & \nodata & \nodata & \nodata   & [10]      \\
                         & \nodata& \nodata & \nodata &  1830  & \nodata & \nodata & \nodata &   840   & \nodata & \nodata & \nodata   & [2]       \\
1118+541                 & \nodata&   1400  & \nodata &  1600  & \nodata & \nodata & \nodata &   450   & \nodata & \nodata & \nodata   & [11]      \\
1136+595                 & \nodata&   1500  &  126    &  1500  & \nodata & \nodata & \nodata &   400   & \nodata & \nodata & \nodata   &           \\
1213+549A                & \nodata& \nodata & \nodata &  1000  &   22    &  1740   & \nodata & \nodata & \nodata & \nodata & \nodata   &           \\
                         & \nodata& \nodata &   41    &   780  & \nodata & \nodata &   23    &   720   &   38    & \nodata & \nodata   & [13]      \\
1258+569                 & \nodata& \nodata &   44    &  1700  &   30    &  2500   &   30    &   800   & \nodata & \nodata & \nodata   &           \\
1315+604                 &  160   &  1700   &   20    &  1300  &   11    &  1800   &   20    &   600   & \nodata & \nodata & \nodata   &           \\
1332+580                 &  250   &  1850   &   36    &  1700  &   16    &  2450   &   10    &   530   &   42    & \nodata & \nodata   &           \\
1340+569                 & \nodata& \nodata &   26    &  1500  & \nodata & \nodata &   10    &   400   & \nodata & \nodata & \nodata   &           \\
1353+564\tablenotemark{a}& \nodata& \nodata & \nodata &  1590  & \nodata & \nodata & \nodata &   520   & \nodata & \nodata & \nodata   & [10]      \\
                         &  241   &  1800   &   41    &  1300  &    9    & \nodata &   71    & \nodata &   40    &   75    & 1300-1780 & [15]      \\
                         & \nodata& \nodata &   57    &  1500  &   20    &  2095   &   85    &   700   & \nodata & \nodata & \nodata   &           \\
1359+536                 &  360   &  1650   &   76    &  1700  &   32    &  2050   &   49    &   500   &   26    & \nodata & \nodata   &           \\
1404+582                 &  150   &  1400   &   38    &  1300  &   17    &   800   &    6    &   530   &   46    & \nodata & \nodata   &           \\
1406+540                 &  100   &  1850   &   28    &  1450  &   10    &  2000   &    8    & \nodata &   27    & \nodata & \nodata   &           \\
1412+538                 &   60   &  1400   &   25    &  1100  &    8    &  1500   &    4    &   400   &   32    & \nodata & \nodata   &           \\
1415+566\tablenotemark{a}&  187   &  2100:  &   17    &   700: & \nodata & \nodata &   11    & \nodata & \nodata & \nodata & \nodata   & [15]      \\
1509+522                 & \nodata& \nodata &  124    &  1800  &   28    &   645   &   62    &  1020   & \nodata & \nodata & \nodata   &           \\
1535+457\tablenotemark{a}& \nodata&  1400L  &  123    &  1680L & \nodata & \nodata & \nodata &   400   & \nodata & \nodata & 1410-1650 & [5]       \\
                         & \nodata& \nodata &  109    &  1480  & \nodata & \nodata &   16    & \nodata &   51    & \nodata & \nodata   & [3]       \\
                         & \nodata& \nodata &  114    & \nodata& \nodata & \nodata &   22    & \nodata & \nodata & \nodata & \nodata   & [16]      \\
1536+498\tablenotemark{a}&  183   &   800   &   29    &  1000  &    2    & \nodata &   18    & \nodata & \nodata &   52    & \nodata   & [15]      \\
1656+578                 &  260   &  2000   &   69    &  1950  &   26    &  2300   &   21    &  500    &   38    & \nodata & \nodata   &           \\

\enddata

\tablenotetext{a}{EW and FWHM of H$\alpha$, H$\beta$, H$\gamma$,
[OIII] and FeII are given in \AA~ in the rest-frame (Stephens
1989, Bade el al. 1995, Moran et al. 1996, Grupe et al.1999).
``L'' means a  Lorenzian fit to the broad component (Veron et al.
2001). In most cases EW and FWHM of H$\alpha$ and H$\gamma$ is the
sum of the blend H$\alpha$+ $[NII]\lambda6584/48$ and
H$\gamma$+$[OIII]\lambda4364$.}

The references in Table \ref{t3} are the same as in Table
\ref{t1}.

In the spectrum of SBS 0921+525, SBS 0933+511, 1055+605, 1135+595,
and SBS 1656+578 very strong and wide HeI$\lambda$5876 emission as
well as $FeII\lambda4924$ are seen. In the spectrum of SBS
0933+511, SBS 0952+552, SBS 1118+541, SBS 1353+564, SBS 1359+536,
SBS 1412+538, [FeVII]$\lambda$6087 and strong HeII$\lambda$4686
are seen.

SBS 1213+549A-- This galaxy was classified as Sy1 by Stepanian et
al. (1991). Martel and Osterbrock (1994) reclassified the object
as NLS1.

SBS 0919+515 is the only object in our sample of NLS1s and in the
sample of AGN of Stephens (1989), for which the FWHM of
$[OIII]\lambda$5007 is greater than FWHM of H$\beta$.

\end{deluxetable}

\clearpage

\begin{deluxetable}{llllll}

\tablecolumns{10} \tabletypesize{\scriptsize} \tablewidth{0pc}
\tablecaption{The Emission line ratios \label{t4}} \tablehead {
\colhead{SBS} & \colhead{[OIII]$\lambda$5007} &
\colhead{H$\alpha$} & \colhead{FeII$\lambda$4570}   &
\colhead{FeII$\lambda$4570}   & \colhead{Ref.}\\
\colhead{Design.} & \colhead{H$\beta$} & \colhead{H$\beta$} &
\colhead{[OIII]} & \colhead{H$\beta$} \\ \colhead{(1)} &
\colhead{(2)} & \colhead{(3)} & \colhead{(4)} & \colhead{(5)} &
\colhead{(6)} }

\startdata

0919+515                  &  0.45   &   2.77  &  3.48   &  1.55   & [2]    \\
0921+525                  & \nodata & \nodata & \nodata &  0.13\tablenotemark{b} & [4,5]  \\
                          &  0.58   & \nodata & (0.24)  &  0.14   & [3]     \\
0924+495                  &  0.25   & \nodata & (4.80)  &   1.2   &         \\
0933+511                  &  $<$3   & \nodata & \nodata & \nodata &         \\
0945+507                  &  0.78   &  5.40   & \nodata & \nodata & [8]     \\
                          &  0.72   & \nodata & (0.83)  &  0.60   & [5]     \\
0952+552\tablenotemark{a} &  0.49   &  2.78   &  0.58   &  0.29   &         \\
1021+561                  &  0.40   & \nodata & (3.43)  &  1.37   &         \\
1022+519\tablenotemark{c} &  0.19   &  3.31   & (4.05)  &  0.77   & [10]    \\
                          &  0.14   & \nodata & (6.57)  &  0.92   & [5]     \\
                          &  0.11   & \nodata & (9.81)  &  1.08   & [3]     \\
1055+605\tablenotemark{c} &  0.30   & \nodata & (2.27)  &  0.68   & [10]    \\
                          &  0.34   &  3.74   &  1.85   &  0.63   & [2]     \\
1118+541                  &  0.5:   & \nodata & \nodata & \nodata & [11]    \\
1136+595                  &  0.30   & \nodata & \nodata & $>1$    &         \\
1213+549A                 & \nodata &  5.10   & \nodata & \nodata & [13]    \\
                          &  0.60   & \nodata & (1.68)  &  1.01   & [12]    \\
                          &  0.45   & \nodata & (2.69)  &  1.21   &         \\
1258+569                  &  0.5:   & \nodata & \nodata & \nodata &         \\
1315+604\tablenotemark{a} &  1.00   &  5.95   & \nodata & \nodata &         \\
1332+580\tablenotemark{a} &  0.25   &  3.60   &  4.32   &  1.08   &         \\
1340+569                  &  1.89   &  3.54   & \nodata & \nodata &         \\
1353+564\tablenotemark{c} &  1.62   & \nodata &  0.30   &  0.52   & [10]    \\
                          &  1.5    & \nodata & \nodata & \nodata & [15]    \\
1359+536\tablenotemark{a} &  0.60   &  2.98   &  0.58   &  0.35   &         \\
1404+582\tablenotemark{a} &  0.23   &  2.82   &  5.65   &  1.30   &         \\
1406+540\tablenotemark{a} &  0.23   &  3.59   &  5.65   &  1.30   &         \\
1412+538\tablenotemark{a} &  0.28   &  3.60   &  5.61   &  1.57   &         \\
1415+566                  &  0.65   & \nodata & \nodata & \nodata & [15]    \\
1509+522\tablenotemark{a} &  0.23   & \nodata & (4.48)  &  1.03   &         \\
1535+547                  &  0.13   & \nodata & (3.54)  &  0.46   & [5]     \\
                          &  0.15   & \nodata & (3.13)  &  0.47   & [3]     \\
1536+498\tablenotemark{a} &  0.58   & \nodata & \nodata & \nodata & [15]    \\
1656+578\tablenotemark{a} &  0.29   &  2.48   &  2.17   &  0.63   &         \\

\enddata

\tablenotetext{a}{The line ratios from Bade et al. (1995) are not
corrected for reddening.} \tablenotetext{b}{The mean value. The
$FeII/H\beta$ range is $0.09-0.16$ according to Veron et al.
(2001).} \tablenotetext{c}{In Grupe et al. (1999), in their table
1 seems like the value of $FeII/H\beta$ and FeII/[OIII] are given,
not the logarithms.}

Data in brackets are calculated from the ratio of $FeII/H\beta$
and [OIII]/H$\beta$.

The references in Table \ref{t4} are the same as in Table
\ref{t1}.

\end{deluxetable}

\clearpage

\begin{deluxetable}{lllllllllllcl}

\tablecolumns{13} \rotate \tabletypesize{\scriptsize}
\tablewidth{0pc} \tablecaption{Optical, ROSAT and radio fluxes and
luminosities for SBS NLS1s \label{t5}} \tablehead{ \colhead{SBS} &
\colhead{$z_o$} & \colhead{B} & \colhead{M} & \colhead{log} &
\multicolumn{2}{c}{ROSAT} & \colhead{log} & \colhead{log} &
\colhead{Radio} & \colhead{log} & \colhead{log} &
\colhead{$\alpha_{ox}$} \\
\colhead{Design.} & & \colhead{$m_{pg}$} & \colhead{B} &
\colhead{L(B)}  & \colhead{cts/s} & \colhead{fx} &
\colhead{f(2keV)} & \colhead{Lx} & \colhead{S(R)} &
\colhead{$L_{R}$}& \colhead{$f_{2keV}/f_{opt}$} & \colhead{} \\
\colhead{(1)} & \colhead{(2)} & \colhead{(3)} & \colhead{ (4)} &
\colhead{(5)} & \colhead{(6)} & \colhead{(7)} & \colhead{(8)} &
\colhead{(9)} & \colhead{(10)} & \colhead{(11)} & \colhead{(12)} &
\colhead{(13)} }

\startdata

0919+515  & 0.161   & 17.33 & -22.6 & 44.56 & 0.418 & 5.21  & -29.24 & 44.76 & \nodata &  \nodata  &  -2.98  & -1.07     \\
0921+525  & 0.0353  & 15.63 & -21.0 & 43.85 & 1.690 &23.01  & -28.60 & 44.11 & 5.16    &   38.60   &  -3.01  & -1.09     \\
0924+495  & 0.1145  & 17.0  & -22.2 & 44.40 &(0.01) &(0.12) &(-30.88)&(42.9) & \nodata &  \nodata  & (-4.74) & $<-1.73$  \\
0933+511  & 0.0553  & 16.5  & -21.1 & 43.94 & 0.054 & 0.65  & -30.15 & 42.96 & \nodata &  \nodata  &  -4.21  & -1.54     \\
0945+507  & 0.0550  & 16.47 & -21.0 & 43.85 & 0.028*& 0.35  & -30.42 & 42.80 & 5.39    &   39.02   &  -4.49  & -1.64     \\
0952+552  & 0.317   & 19    & -22.6 & 44.56 & 0.037*& 0.46  & -30.30 & 44.42 & \nodata &   \nodata &  -3.36  & -1.22     \\
1021+561  & 0.197   & 18.02 & -22.4 & 44.49 & 0.117 & 1.09  & -29.92 & 44.34 & \nodata &   \nodata &  -3.37  & -1.22     \\
1022+519  & 0.045   & 15.81 & -21.4 & 44.08 & 1.750 &20.10  & -28.66 & 44.35 & 1.08    &   38.14   &  -3.00  & -1.08     \\
1055+605  & 0.149   & 17.2  & -22.6 & 44.56 & 0.389 & 3.22  & -29.45 & 44.55 & \nodata &   \nodata &  -3.23  & -1.17     \\
1118+541  & 0.1043  & 16.41 & -22.6 & 44.56 & 0.265 & 2.74  & -29.52 & 44.15 & 2.98    &   39.33   &  -3.62  & -1.31     \\
1136+595  & 0.1138  & 17.0  & -22.2 & 44.40 &(0.01) &(0.12) &(-30.88)&(42.9) & \nodata &   \nodata & (-4.74) & $<-1.73$  \\
1213+549A & 0.1505  & 16.91 & -22.9 & 44.70 & 0.090 & 1.19  & -29.88 & 44.12 & 2.12    &   39.52   &  -3.78  & -1.37     \\
1258+569  & 0.0719  & 17    & -21.3 & 43.97 & 0.066 & 0.82  & -30.05 & 43.29 & \nodata &   \nodata &  -3.91  & -1.42     \\
1315+604  & 0.1372  & 17.5  & -22.1 & 44.36 & 0.197 & 2.65  & -29.54 & 44.38 & 2.7\tablenotemark{a}& 39.54\tablenotemark{a}& -3.20 & -1.16 \\
1332+580  & 0.124   & 17.5  & -21.9 & 44.31 & 0.126 & 1.58  & -29.76 & 44.07 & \nodata &   \nodata &  -3.41  & -1.24     \\
1340+569  & 0.0396  & 16.98 & -19.9 & 43.45 & 0.122 & 1.31  & -29.84 & 42.96 & \nodata &   \nodata &  -3.71  & -1.35     \\
1353+564  & 0.1223  & 17    & -22.4 & 44.46 & 0.546 & 6.19  & -29.17 & 44.66 & 6.14    &   39.79   &  -3.03  & -1.10     \\
1359+536  & 0.175   & 18.5  & -21.7 & 44.17 & 0.035*& 0.44  & -30.32 & 43.83 & \nodata &   \nodata &  -3.58  & -1.30     \\
1404+582  & 0.125   & 17.5  & -21.9 & 44.31 & 0.104 & 1.30  & -29.85 & 43.99 & \nodata &   \nodata &  -3.50  & -1.28     \\
1406+540  & 0.173   & 18.5  & -21.7 & 44.17 & 0.027*& 0.34  & -30.43 & 43.71 & \nodata &   \nodata &  -3.69  & -1.34     \\
1412+538  & 0.164   & 17.5  & -22.5 & 44.53 & 0.066 & 0.75  & -30.08 & 44.01 & \nodata &   \nodata &  -3.74  & -1.36     \\
1415+566  & 0.150   & 17.09 & -22.7 & 44.62 & 0.222 & 2.83  & -29.51 & 44.50 & \nodata &   \nodata &  -3.33  & -1.21     \\
1509+522  & 0.210   & 17.65 & -22.9 & 44.70 &(0.01) &(0.12) &(-30.88)&(43.4) & \nodata &   \nodata & (-4.49) & $<-1.64$  \\
1535+547  & 0.0397  & 15.21 & -21.6 & 44.17 &(0.01) &(0.12) &(-30.88)&(41.9) & 1.25    &   38.09   & (-5.48) & $<-2.01$  \\
1536+498  & 0.280   & 18.3  & -22.9 & 44.70 & 0.387 & 5.25  & -29.24 & 45.36 & 1.63    &   40.00   &  -2.58  & -0.93     \\
1656+578  & 0.198   & 18.5  & -21.9 & 44.31 & 0.053 & 0.91  & -30.00 & 44.27 & \nodata &   \nodata &  -3.26  & -1.18     \\

\enddata

\tablenotetext{a}{NVSS, Bauer et al. (2000)}

*) RASS faint source catalogue (Voges et al. 2000). Data in
brackets are upper limit estimates.

\end{deluxetable}

\clearpage

\begin{deluxetable}{lllllllllllllll}

\tablecolumns{15} \tabletypesize{\scriptsize} \tablewidth{0pc}
\tablecaption{IRAS fluxes and luminosity for Mkn 124 \label{t6}}
\tablehead{ \colhead{SBS} & \multicolumn{8}{c}{IRAS} &
\multicolumn{6}{c}{log}\\
\cline{2-8} \cline{10-15} \colhead{Design.} & \colhead{12} &
\colhead{Q} & \colhead{25} & \colhead{Q} & \colhead{60} &
\colhead{Q} & \colhead{100} & \colhead{Q} & \colhead{12/25} &
\colhead{25/60} & \colhead{60/100} & \colhead{$F_{FIR}$} &
\colhead{$L_{FIR}/L_{\odot}$} & \colhead{$L_{FIR}$}
\\\colhead{(1)} & \colhead{(2)} & \colhead{(3)} & \colhead{
(4)} & \colhead{(5)} & \colhead{(6)} & \colhead{(7)} &
\colhead{(8)} & \colhead{(9)} & \colhead{(10)} & \colhead{(11)} &
\colhead{(12)} & \colhead{(13)} & \colhead{(14)} & \colhead{(15)}}

\startdata

0945+507 & 0.119 & 2 & 0.266 & 3 & 0.683 & 3 & 0.78 & 2 & -0.351 & -0.409 & -0.058 & -13.49 & 11.07 & 44.65 \\

\enddata

\end{deluxetable}

\begin{deluxetable}{lllll}

\tablecolumns{13} \tabletypesize{\small} \tablewidth{0pc}
\tablecaption{Correlations for SBS NLS1s \label{t7}} \tablehead{
\colhead{Index} & \colhead{Variables}
& \colhead{N} & \colhead{P} & \colhead{R}  \\
\colhead{(1)} & \colhead{(2)} & \colhead{(3)} & \colhead{(4)} &
\colhead{(5)} }

\startdata

1  & $EW_{H\beta}$ -- $FWHM_{H\beta}$ & 24  &  0.00521   &  0.55155    \\
2  & $FeII/H\beta$ -- $EW_{H\beta}$   & 17  &  0.01245   & -0.59116    \\
3  & $FeII/H\beta$ -- $FWHM_{H\beta}$ & 17* &  0.00039   & -0.76071    \\
4  & $FeII/[OIII]$ -- $FWHM_{H\beta}$ & 17* &  0.00021   & -0.78144    \\
5  & $FeII/[OIII]$ -- [OIII]/H$\beta$ & 18  &  0.00014   & -0.77805    \\
6  & $FeII/H\beta$ -- $FeII/[OIII]$   & 19  & $<0.00001$ &  0.80386    \\
7  & $logL_{R}$    -- $EW_{H\beta}$   &  8  &  0.04455   & -0.71875    \\
8  & $logL_{R}$    -- $logL_{B}$      &  9  &  0.03095   &  0.71335    \\
9  & $logL_{R}$    -- [OIII]/H$\beta$ &  9  &  0.04242   &  0.68339    \\
10 & $logL_{R}$    -- $FWHM_{[OIII]}$ &  9  &  0.03853   &  0.83514    \\
11 & $logL_{R}/L_{x}$--$logL_{R}$     &  9  &  0.00081   &  0.90445    \\
12  & $logL_{x}$     --$logL_{B}$      & 23  & $<0.00001$ &  0.77880    \\
13  & $\alpha_{ox}$  --$logL_{x}$      & 23  & $<0.00001$ &  0.87029    \\

\enddata

*) Two discrepant points were discarded for these comparisons.
Including them yield  N=19, P=0.00741 and 0.08207, and R=-0.59332
and -0.40901 for comparisons 3 and 4, respectively.

\end{deluxetable}

\clearpage

\end{document}